\documentclass[aps,showpacs,superscriptaddress,preprint]{revtex4-2}

\usepackage{amssymb,amsmath,amsfonts,amsthm,latexsym,graphicx,epsfig,fancybox}
\usepackage{titlesec}
\usepackage{float}
\usepackage{empheq}
\usepackage{epstopdf}
\usepackage{xspace,color,xcolor}
\usepackage{bm}
\usepackage{booktabs}

\usepackage [latin1]{inputenc}

\titleformat{\section}{\large\bfseries}{\thesection}{1em}{}

\newcommand{\bea}{\begin{eqnarray}}
\newcommand{\ena}{\end{eqnarray}}
\newcommand{\be}{\begin{equation}}
\newcommand{\en}{\end{equation}}
\newcommand{\nn}{\nonumber\\}
\newcommand{\ed}{\end{document}}

\newcommand{\Tr}{\mbox{\rm{tr}}}

\newcommand{\Br}{\ensuremath{\mathcal{B}}\xspace}
\newcommand{\Jpsi}{\ensuremath{J\!/\!\psi}\xspace}

\begin{document}

\title{Revisiting radiative transitions of charmonium states\\
  in covariant confined quark model}
\author{Aidos Issadykov}
\affiliation{Bogoliubov Laboratory of Theoretical Physics, JINR,
  Joliot-Curie 6, 141980 Dubna, Russia}
\affiliation{The Institute of Nuclear Physics, Ministry of Energy of
the Republic of Kazakhstan, Almaty, 050032 Kazakhstan}
\author{Mikhail A. Ivanov}
\affiliation{Bogoliubov Laboratory of Theoretical Physics, JINR,
  Joliot-Curie 6, 141980 Dubna, Russia}
\author{Dang-Khoa N.  Nguyen}
\affiliation{Department of Physics, HCMC University of Technology
	and Education, Vo Van Ngan 1, 700000 Ho Chi Minh City, Vietnam}
\author{Chien-Thang  Tran}
\email{thangtc@hcmute.edu.vn}
\affiliation{Department of Physics, HCMC University of Technology
  and Education, Vo Van Ngan 1, 700000 Ho Chi Minh City, Vietnam}
\author{Akmaral Tyulemissova}
\affiliation{Bogoliubov Laboratory of Theoretical Physics, JINR,
  Joliot-Curie 6, 141980 Dubna, Russia}
\affiliation{The Institute of Nuclear Physics, Ministry of Energy of
the Republic of Kazakhstan, Almaty, 050032 Kazakhstan}
\author{Zhomart Tyulemissov}
\affiliation{Bogoliubov Laboratory of Theoretical Physics, JINR,
  Joliot-Curie 6, 141980 Dubna, Russia}
\affiliation{The Institute of Nuclear Physics, Ministry of Energy of
the Republic of Kazakhstan, Almaty, 050032 Kazakhstan}
\affiliation{Al-Farabi Kazakh National University, Almaty, Kazakhstan}
\begin{abstract}
We calculate the amplitudes and branching fractions of $\chi_{cJ}$, $h_c$, $J/\psi$, and $\psi(2S)$ radiative decays in the framework of the covariant confined quark model. First, we explicitly show that all calculated amplitudes are gauge invariant. Next, we use experimental data for three decay channels $J/\psi\to \eta_c\gamma$, $\chi_{c0}\to J/\psi\gamma$, and $\psi(2S)\to \chi_{c0}\gamma$ to fit the model parameters including the charm quark mass and the meson size parameter. Finally, we use the obtained parameters to predict the branching fractions of the decays $h_c\to\eta_c\gamma$, $\chi_{c1(c2)}\to J/\psi\gamma$, and $\psi(2S)\to\chi_{c1(c2)}\gamma$. Our predictions agree well with experimental data. 
\end{abstract}
\maketitle
\newpage

\section{Introduction}
\label{sec:intro}

Charmonium states with masses below the $D\bar D$  threshold attract much attention
from both theorists and experimentalists. The main reason is that these states
have  dominant radiative decays with narrow decay widths.
To date, radiative decays of the $\eta_c(1S)$, $J/\psi(1S)$, $\chi_{c0,c1,c2}(1P)$,
$h_c(1P)$, $\eta_c(2S)$, and $\psi(2S)$ charmonium states 
have been observed and studied experimentally.
The most comprehensive overview of the field up to 2011 was presented
in Ref.~\cite{Brambilla:2010cs}.

Branching fractions for radiative transitions from the $\psi(2S)$ to the $\chi_{c0,c1,c2}$ states
and also to the $\eta_c$ were presented in Ref.~\cite{Gaiser:1985ix}. These results were obtained from a detailed study using the Crystal Ball detector at SPEAR (SLAC). Additionally, the resonance parameters of the $\chi_{c0}$ were measured at the Fermilab Antiproton
Accumulator~\cite{FermilabE835:2002pkx}. 
%
The CLEO collaboration, using the CLEO~III detector at the Cornell Electron Storage Ring, has performed several key studies on charmonium states. In particular, they investigated photon transitions in $\psi(2S)$ decays to $\chi_{cJ}(1P)$ and $\eta_c(1S)$ states~\cite{CLEO:2004cbu}. Their work also includes the measurement of the branching fractions for the $\chi_{cJ}\to\gamma J/\psi$ and $\psi(2S)\to\textit{light hadrons}$ decays~\cite{CLEO:2005efp}, as well as for $\psi(2S)\to J/\psi$ transitions with greater precision than previously achieved~\cite{CLEO:2008kwj}. Furthermore, the CLEO collaboration observed the $h_c(^1P_1)$ state of charmonium in the reaction $\psi(2S)\to \pi^0 h_c \to (\gamma\gamma)(\gamma\eta_c)$~\cite{CLEO:2005vqq} and later reported a precise measurement of its mass~\cite{CLEO:2008ero}. In a separate study, they presented the most precise measurements of the radiative decays $\psi(2S)\to \eta_c\gamma$ and $J/\psi\to\eta_c\gamma$~\cite{CLEO:2008pln}.


Using photon conversions to $e^+e^-$ pairs, the energy spectrum of inclusive
photons from $\psi(2S)$ radiative decays was measured by BESII at the
Beijing Electron-Positron Collider~\cite{BES:2005bmx}. The $\chi_{cJ}(1P)$ states ($J=0,1,2$) were clearly observed  and their masses  were determined.
The process $\psi(3686) \to \pi^0h_c,  h_c\to\gamma\eta_c$ was studied
in Ref.~\cite{BESIII:2012urf}. The mass and decay width of the $h_c(^1P_1)$ state were determined
by simultaneously fitting distributions of the $\pi^0$ recoil mass for 16 exclusive $\eta_c$ decay modes.
Using a sample of 106 million $\psi(3686)$ decays, the branching fractions
$\psi(3686)\to\gamma\chi_{c0}$,   $\psi(3686)\to\gamma\chi_{c1}$, and
 $\psi(3686)\to\gamma\chi_{c2}$
were measured with improved precision~\cite{BESIII:2017gcu}.
First measurements of the absolute branching fractions
$\Br(\psi'\to\pi^0 h_c)  = (8.4 \pm 1.3 \pm 1.0)\times 10^{-4}$ and
$\Br(h_c\to\gamma\eta_c) = (54.3 \pm 6.7 \pm 5.2)\%$ were also reported by the BESIII collaboration~\cite{BESIII:2010gid}.
Using 448 million $\psi(2S)$ events, the spin-singlet $P$-wave charmonium state $h_c(1^1P_1)$
was studied via the $\psi(2S)\to\pi^0 h_c$ decay followed by the $h_c\to\gamma\eta_c$
transition~\cite{BESIII:2022tfo}.
The branching fractions were measured to be
$\Br_{\rm Inc}(\psi(2S)\to\pi^0 h_c)\times\Br_{\rm Tag}(h_c\to\gamma\eta_c)
= (4.22^{+0.27}_{-0.26}\pm0.19)\times 10^{-4}$, 
$\Br_{\rm Inc}(\psi(2S)\to\pi^0 h_c)= (7.32 \pm 0.34 \pm 0.41)\times 10^{-4}$,  and
$\Br_{\rm Tag}(h_c\to\gamma\eta_c) = (57.66^{+3.62}_{-3.50} \pm 0.58)\%$,
where the uncertainties are statistical and systematic, respectively.
The $h_c(1^1P_1)$ mass and width were determined to be
$M = (3525.32 \pm 0.06 \pm 0.15)$~MeV and
$\Gamma  = (0.78^{+0.27}_{-0.24 }\pm 0.12)$~MeV.
The Particle Data Group (PDG) combined their best value of $\Br(\psi(2S)\to  h_c(1P) \pi^0)
= (7.4\pm0.5)\times 10^{-4}$~\cite{ParticleDataGroup:2024cfk} with the result 
\[
\Br(h_c \rightarrow \gamma \eta_c)\times \Br(\psi(2S)\to h_c(1P)
\pi^0) = (4.22^{+0.27}_{-0.26} \pm 0.19 )\times 10^{-4}
\]
reported by BESIII~\cite{BESIII:2022tfo} and found 
\[
\Br(h_c \rightarrow \gamma \eta_c) =
\left\{\begin{array}{l}
(60\pm 4)\%\qquad\text{FIT}\\
(57\pm 5)\%\qquad\text{AVERAGE}
\end{array}\right. 
\]


Radiative decays in  charmonium play an important role in the understanding
of its structure and can serve a testing ground for a number of theories and
models. In Ref.~\cite{Barnes:2005pb},  results for the spectrum and radiative partial
widths were presented. They have been evaluated using two models, the relativized
Godfrey-Isgur model and a nonrelativistic potential model. The electromagnetic
transitions were evaluated using Coulomb plus linear plus smeared hyperfine wavefunctions,
both in the nonrelativistic potential model and in the Godfrey-Isgur model.
The available information on quarkonia and their transitions
has been reviewed in Ref.~\cite{Eichten:2007qx} and  theoretical implications
have been  discussed. 
Topics in the description of the properties of charmonium states were
reviewed in Ref.~\cite{Voloshin:2007dx} with an emphasis on specific theoretical ideas
and methods of relating those properties to the underlying theory of Quantum Chromodynamics.
The masses, electromagnetic decays, and E1 transitions of
charmonium states were calculated in the screened potential model~\cite{Li:2009zu}.
Study of the mass spectrum and electromagnetic processes of charmonium system
was carried out  in Ref.~\cite{Cao:2012du}
with the spin-dependent potentials fully taken into account in the solution
of the Schr\"odinger equation.
The charmonium spectrum was calculated with two nonrelativistic quark models,
the linear potential model and the screened potential model in Ref.~\cite{Deng:2016stx}.
Using the obtained wavefunctions,  the electromagnetic transitions of charmonium states
were evaluated.

The transition form factors of various multipolarities between the lightest few charmonium states were computed for the first time within lattice QCD (LQCD) in Ref.~\cite{Dudek:2006ej}. In a subsequent publication~\cite{Dudek:2009kk}, this method was applied to compute radiative transition rates involving excited charmonium states,
states of high spin, and exotics.
Charmonium radiative transitions  $J/\psi\to\eta_c\gamma$, $\chi_{c0}\to J/\psi \gamma$, and $h_c\to \eta_c\gamma$  were calculated  using $N_f = 2$ twisted mass lattice
 QCD gauge configurations~\cite{Chen:2011kpa}. 
Hadronic matrix elements relevant to the $h_c\to \eta_c\gamma$  and $h_b\to \eta_b\gamma$ decays were computed within LQCD  by using the gauge configurations produced by the Extended Twisted Mass Collaboration with $N_f = 2 + 1 + 1$ dynamical Wilson-Clover
twisted mass fermions in Ref.~ \cite{Becirevic:2025ocx}.

Dispersion sum rules have also been applied to the calculation of radiative transition amplitudes in quarkonium, see e.g., Ref.~\cite{Khodjamirian:1979fa}. 
Within the QCD sum rule approach, the radiative decay $J/\psi \to \eta_c \gamma$
was analyzed~\cite{Beilin:1984pf} taking into account both nonperturbative
and perturbative corrections.
The radiative decays of heavy $\bar Q Q$ state were studied in Ref.~\cite{DeFazio:2008xq}
using an effective Lagrangian approach which exploits spin symmetry for such states.
The radiative transitions among the vector and scalar heavy quarkonium states
were studied in Ref.~\cite{Wang:2012ph} within the covariant light-front quark model.
It was observed that the radiative decay widths are sensitive
to the constituent quark masses and the shape parameters of the wave-functions.

In the paper~\cite{Brambilla:2005zw} magnetic dipole (M1) transitions
between two quarkonia were studied in the framework of
nonrelativistic effective field theories of QCD.
This study was extended in Ref.~\cite{Brambilla:2012be} to explore electric
dipole (E1) transitions in heavy quarkonium. 
The determination of heavy quarkonium magnetic dipole transitions
were updated and improved in Ref.~\cite{Pineda:2013lta}
using the potential nonrelativistic QCD.

The mass spectra and electromagnetic decay rates of charmonium, bottomonium,
and $B_c$ mesons were comprehensively investigated in the relativistic quark
model~\cite{Ebert:2002pp}.
Theoretical description of charmonium radiative decays 
was provided in Ref.~\cite{Bruschini:2019vff} in the framework of 
a simple quark potential model beyond the so-called standard
$p/m$ approximation.
Radiative decays of $0^{++}$ and $1^{+-}$ heavy mesons
were studied in Ref.~\cite{Ke:2013zs} within the light front quark model.
In Ref.~\cite{Guo:2014zva}, radiative transitions between conventional
charmonium states and from the lowest multiplet of $c\bar c$
hybrids to charmonium mesons were studied by considering the nonrelativistic limit of the QCD Hamiltonian in the Coulomb gauge.



In this study, we aim to  evaluate the amplitudes and branching fractionss of radiative
decays of charmonium states $\chi_{c0,c1,c2}\to J/\psi \gamma$,
$\psi(2S)\to\chi_{c0,c1,c2}\gamma$,  $h_c\to \eta_c\gamma$, and $J/\psi\to\eta_c\gamma$
in the framework of the covariant confined quark model (CCQM) previously developed by us. Keeping in mind that some of the radiative decays
mentioned above have been evaluated in the framework
of the CCQM before~\cite{Ganbold:2021nvj}, one has to emphasize  how this work differs from the previous one.
First, we accept here the convential method of electromagnetic gauging of
the nonlocal Lagrangian which was suggested in Refs.~\cite{Mandelstam:1962mi,Terning:1991yt}
and intensively used in our previous papers~\cite{Ivanov:1996pz,Ivanov:1998wj,Ivanov:1999bk,Branz:2009cd,Branz:2010pq,Dubnicka:2011mm,Gutsche:2012ze,Dubnicka:2024geu}.
Second, we add three more modes $\psi(2S)\to\chi_{c0,1,2}\gamma$ in the present study.
Finally, we provide a new fit of the charm quark mass and meson size parameters, and calculate the propagation of errors from the fit to the final theoretical predictions. 

The paper is organized as follows. In Sec.~\ref{sec:formalism} we present the theoretical framework of the study. Here we briefly discuss how charmonium states are described within the CCQM and how the gauging process of nonlocal quark current proceeds. We show that all calculated amplitudes are gauge invariant. In Sec.~\ref{sec:Gam} we obtain all expressions for the decay width of radiative charmonium decays in terms of helicity amplitudes. The fitting process and numerical results are presented in Sec.~\ref{sec:result}. A brief conclusion is given in Sec.~\ref{sec:sum}.   


\section{Theoretical framework}
\label{sec:formalism}

\subsection{Charmonium in covariant confined quark model
}

In this study, we consider a large set of charmonium states of different quantum numbers and titles. Therefore, for the convenience of the reader, we list in Table~\ref{tab:states} a brief summary for the charmonium states treated in this paper. Their masses and widths are taken from the 
PDG~\cite{ParticleDataGroup:2024cfk}.
\begin{table}[htbp]
\caption{The charmonium states $^{2S+1}L_{\,J}$.
We use the notation $\stackrel{\leftrightarrow}{\partial}=
\stackrel{\rightarrow}{\partial}-\stackrel{\leftarrow}{\partial}$.
The numerical values of masses and decay widths are taken from PDG~\cite{ParticleDataGroup:2024cfk}.}
\label{tab:states}
\def\arraystretch{1.5}
\begin{ruledtabular}
		\begin{tabular}{ccccc}
			Quantum number & Title & Quark current & Mass (MeV) & Width (MeV)\\
			\hline
			$J^{PC}=0^{-+}$  & $^1S_0=\eta_c$ & $\bar q\, i\gamma^5\, q $ & 2984.1(4) & 30.5(5)  \\		
			$J^{PC}=1^{--}$ & $^3S_1=J/\psi$ & $\bar q\,\gamma^\mu\, q $ & 3096.900(6) & 0.0926(17) \\			
			$J^{PC}=0^{++}$ & $^3P_0=\chi_{c0}$ & $\bar q\, q $ & 3414.71(30) & 10.5(8)   \\		
			$J^{PC}=1^{++}$ & $^3P_1=\chi_{c1}$ & $\bar q\,\gamma^\mu\gamma^5\,q $ & 3510.67(5) & 0.84(4)   \\		
			$J^{PC}=1^{+-}$ & $^1P_1=h_c(1P)$ & $\bar q\, \stackrel{\leftrightarrow}{\partial}^{\,\mu} \gamma^5\, q $ &
			3525.37(14) & 0.78(28)  \\
			$J^{PC}=2^{++}$  & $^3P_2=\chi_{c2}$ &
			$(i/2)\,\bar q\, \left(\gamma^\mu \stackrel{\leftrightarrow}{\partial}^{\,\nu}
			+\gamma^\nu \stackrel{\leftrightarrow}{\partial}^{\,\mu}\right)\,q $ &
			3556.17(7) & 1.98(9)  \\		
			$J^{PC}=1^{--}$  & $^3S_1=\psi(2S)$ &  $\bar q\,\gamma^\mu\, q $ & 3686.097(11) & 0.293(9)
		\end{tabular}
\end{ruledtabular}
\end{table}
The CCQM is based on a relativistic
Lagrangian describing the interaction of a hadron with its constituent quarks. 
The charmonium is described by a field $\phi_{cc}(x)$
which couples to an interpolating quark current $J_{cc}(x)$.
The  interaction Lagrangian reads
\be
{\cal L}_{\rm int}(x) = g_{cc}\,\phi_{cc}(x)\cdot J_{cc}(x).
\label{eq:Lag}
\en
The quark current $J_{cc}(x)$ is a nonlocal generalization
of the quark currents shown in Table~\ref{tab:states}:
\bea
J_{cc}(x) &=& \int\!\!\!\int\! dx_1dx_2\, F_{\rm cc}(x,x_1,x_2)\cdot
\bar c(x_1)\,\Gamma_{cc}\, c(x_2),
\label{eq:quark_cur}\\[2ex]
F_{cc}(x,x_1,x_2)&=&\delta(x-w\,x_1-w\,x_2)
\,\Phi_{cc}\left((x_1-x_2)^2\right), \qquad (w=1/2).
\nonumber
\ena
The Fourier transform of the translationally invariant vertex function
$\Phi_{cc}\left((x_1-x_2)^{2}\right)$ in momentum space
is required to fall off in the Euclidean region in order to ensure the ultraviolet convergence of the loop integrals. We use a simple Gaussian form written as
\be
\widetilde{\Phi}_{cc}\left(-p^2\right)
= \exp\left( s_{cc} \!\cdot\! p^2  \right), \qquad   s_{cc} \equiv
1 / \Lambda_{cc}^2 \,,
\label{eq:vertex}
\en
where $\Lambda_{cc}$ is an adjustable charmonium size-related parameter of
the CCQM. The choice of this parameter will be discussed later on.
One has to note that in the case of the radial excitation $\psi(2S)$, we use an alternative form for the vertex function~\cite{Dubnicka:2024geu}
\be
\widetilde{\Phi}_{2S}\left(-p^2\right)
= (1 + c_1 s_{2S} p^2 )\exp\left( s_{2S} \!\cdot\! p^2  \right), \qquad   s_{2S} \equiv
1 / \Lambda_{2S}^2 \,.
\label{eq:vertex-2S}
\en
The coefficient $c_1$ is determined from the orthogonality condition
which was suggested and discussed in our paper~\cite{Dubnicka:2024geu}.

The coupling constant $g_{cc}$ in Eq.~(\ref{eq:Lag}) is calculated from the so-called compositeness condition,
which is expressed in terms of the derivative of
the scalar part of the charmonium mass operator:
\be
Z_{cc} =  1-g_{cc}^{2}\widetilde{\Pi}_{cc}^\prime(m^2_{cc})=0.
\label{eq:coupling}
\en
The calculation details can be found in our previous papers~\cite{Ivanov:2005fd,Ganbold:2021nvj,Tran:2023hrn}.

\subsection{Gauging of nonlocal quark current}

The gauge invariant interaction of a bound quark state with the 
electromagnetic field has been described in some detail 
in Ref.~\cite{Ivanov:1996pz}. For comprehensive
purposes we recall some of the key points of the gauging process.

In order to guarantee gauge invariance of the nonlocal strong interaction 
Lagrangian, one multiplies each quark field $q(x_i)$ 
by a gauge field exponentional according to
\bea 
\label{eq:gauging}
 q(x_i) &\to& Q(x_i) = e^{-ie_q I(x_i,x,P)}\,  q(x_i),\qquad  
I(x_i,x,P) = \int\limits_x^{x_i} dz_\mu A^\mu(z),
\nn
\bar q(x_i) &\to&  \bar Q(x_i)  = e^{ie_q I(x_i,x,P)}\,  \bar q(x_i),
\ena 
where $P$ is the path taken from $x$ to $x_i$.
It is readily seen that the neutral nonlocal quark current defined
by
\be
J^{\,\rm em}(x) = \int\!\!\!\int\! dx_1dx_2\, \delta(x-w\,x_1-w\,x_2)
\,\Phi\left((x_1-x_2)^2\right) \bar Q(x_1)\,\Gamma\, Q(x_2),
\quad (w=1/2), 
\label{eq:quark_cur-em}
\en
is invariant under the local gauge transformations
\bea 
q(x_i) &\to& e^{ie_q f(x_i)} q(x_i),    \qquad
\bar q(x_i) \to e^{-ie_q f(x_i)} \bar q(x_i),
\nn  
A^\mu(z)&\to& A^\mu(z)+\partial^\mu f(z), \quad\text{so that} \quad
I(x_i,x,P)\to I(x_i,x,P) + f(x_i) - f(x),
\label{eq:gauge-group}
\ena
if the matix $\Gamma$ has no derivative.

The first term of the electromagnetic interaction
arises when one expands the gauge exponential
in powers of $I(x_i,x,P)$.
Superficially, the results appear to depend on the path $P$.
However, one needs to know only derivatives of the path
integrals when doing the perturbative expansion.
One can make use of the formalism developed in~\cite{Mandelstam:1962mi,Terning:1991yt}
which is based on the path-independent definition of the derivative of 
$I(x,y,P)$:
\be\label{eq:path}
\frac{\partial}{\partial x^\mu} I(x,y,P) = A_\mu(x)
\en
which states that the derivative of the path integral $I(x,y,P)$ does 
not depend on the path $P$ originally used in the definition. 
It is easy to check that such procedure of gauging the free quark lagrangian 
leads to the standard form of $e_q\, \bar q(x)\,\hat A(x) q(x)$.

Expanding the gauged quark current given by Eq.~(\ref{eq:quark_cur-em})
up to the first order in $A^{\mu}$ one obtains 
\bea
J^{\rm em}(x) &=& \int\!\!\!\int\! dx_1dx_2\, \delta(x-w\,x_1-w\,x_2)
\,\Phi\left((x_1-x_2)^2\right)\,\bar q(x_1)\,\Gamma\, q(x_2)
\nn
&\times& \left(1 +ie_q I(x_1,x_2)+\mathcal{O}(e_q^2)\right).
\label{eq:quark_cur-em-1}
\ena
If a quark current contains derivatives, e.g. in the case of the $h_c$ and
$\chi_{c2}$ charmonium states (see Table~\ref{tab:states}), extra terms arise in Eq.~(\ref{eq:quark_cur-em-1}). For instance,
for $h_c$ one has
\bea
J^{\rm em}_{h_c}(x) &=& \int\!\!\!\int\! dx_1dx_2\, \delta(x-w\,x_1-w\,x_2)
\,\Phi_{h_c}\left((x_1-x_2)^2\right)\,
\nn
&\times&
\Big\{
\big[\bar q(x_1)\,\gamma_5\,\partial^\mu_{x_2} q(x_2)
      -\partial^\mu_{x_1}\bar q(x_1)\,\gamma_5\,q(x_2)\big] 
\,\big(1 +ie_q I(x_1,x_2)\big)
\nn
&-& i e_q \big[A^\mu(x_1) + A^\mu(x_2)\big]
\bar q(x_1)\,\gamma_5\, q(x_2)
\Big\} + \mathcal{O}(e_q^2).
\label{eq:quark_cur-em-2}
\ena

The evaluation of Feynman diagrams involving the strong quark vertex with
emitting photon leads to a typical integral
\be
R(x;k_1,k_2) =  \int\!\!\!\int\! dx_1dx_2\, \delta(x-w\,x_1-w\,x_2)
\,\Phi\left((x_1-x_2)^2\right) I(x_1,x_2)\,e^{ik_1x_1-ik_2x_2},
\label{eq:int-R}
\en
which is calculated by replacing
the integration variables $x_1=z-w\rho$ and $x_2=z+w\rho$
where $w=1/2$. One has
\[
R(x;k_1,k_2) =e^{i(k_1-k_2)x} \int\! d\rho\, \Phi(\rho^2)\,
I\big(x-w\rho\,,\,x+w\rho\big)\,e^{-ik\rho},
\quad\text{where}\quad k=\tfrac12(k_1+k_2).
\]
In the next step, one can use the transformation of the vertex function
given by
\[
\Phi(\rho^2) = \int\frac{d^4\ell}{(2\pi)^4} e^{-i\ell\rho}\widetilde\Phi(-\ell^2)
=\widetilde\Phi(\partial_\rho^2) \delta^{(4)}(\rho).
\]
Finally, we are coming to the expression which has been calculated
in Refs.~\cite{Ivanov:1996pz,Branz:2009cd,Branz:2010pq,Dubnicka:2024geu}.
Assuming that $A_\alpha(x)=\epsilon^\ast_\alpha e^{iqx}$ one has
\bea
R(x,k_1,k_2) &=&  e^{i(k_1-k_2)x}
\int\! d^4\rho\,\delta^{(4)}(\rho)\,\widetilde\Phi(\partial_\rho^2)\,
I\big(x-w\rho\,,\,x+w\rho\big)\,e^{-ik\rho}
\nn
&=& i\,\epsilon^\ast_\alpha\,e^{i(k_1-k_2+q)x}
\int\limits_0^1\! d\tau\,
\Big\{\widetilde\Phi'\big(-z^+_\tau\big)\,(k + w^2 q)^\alpha
     +\widetilde\Phi'\big(-z^-_\tau\big)\,(k - w^2 q)^\alpha\Big\},
     \label{eq:int-R-final}\\
 z^\pm_\tau &=& (k \pm w\,q)^2\,\tau + k^2\,(1-\tau)\,.    
     \nonumber
\ena

\subsection{Radiative decays \boldmath{$(\bar cc)_1\to (\bar cc)_2+\gamma$}:
  Feynman diagrams and loop integrals}

\begin{figure}[H]
\begin{centering}
  \includegraphics[width=0.8\textwidth]{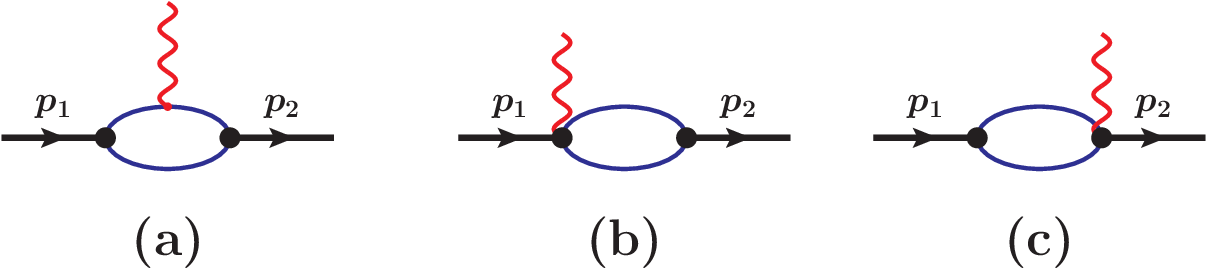}
\par\end{centering}
\caption{Diagrams describing the $(\bar cc_1)\to (\bar cc)_2 + \gamma$
  transition.}
\label{fig:diagrams}
\end{figure}

In this paper we will consider eight modes of charmonium
radiative decays:
\begin{eqnarray*}
&&  
(a)\,\,\, \chi_{cJ}\to J/\psi + \gamma, \quad\,\,\, (b)\,\,\, 
  \psi(2S)\to \chi_{cJ} + \gamma\,, \quad\text{where}\,\,\, J=0,1,2,
\nn  
&& (c) \,\,\, h_c\to \eta_c +\gamma \quad\text{and}\quad
J/\psi\to \eta_c + \gamma\,.
\end{eqnarray*}
The relevant Feynman diagrams are shown in Fig.~\ref{fig:diagrams}.
Due to the gauging technique described above, there are three topologies, first,
the diagram with photon emitted from quark line, and second, two diagrams
with photons emitted from the nonlocal vertices.

We start with the group of the 
$\chi_{cJ}(p_1)\to J/\psi(p_2) + \gamma(q)$ ($J=0,1,2$) decays.
The invariant matrix element describing these decays are written as
\be
M_{\chi_{cJ}\to J/\psi+ \gamma} = 6\,g_{\chi_{cJ}}\,g_{J/\psi}\,e_q\,
\epsilon^\ast_{2\,\beta}(p_2)\,\epsilon^\ast_{\gamma\,\alpha}(q)\,
M^{\beta\alpha}\,,
\label{eq:inv_ampl}
\en
where the amplitude $M^{\beta\alpha}$ is written via  loop integrals corresponding to the Feynman diagrams. One has
\bea
M^{\beta\alpha} &=& 
M^{\beta\alpha}_{\triangle\,a} + M^{\beta\alpha}_{\bigcirc\,b} +  M^{\beta\alpha}_{\bigcirc\,c},
\nn[2ex]
M^{\beta\alpha}_{\triangle\,a} &=& 
-\int\frac{d^4\ell}{(2\pi)^4i} \widetilde\Phi_{cJ}\big(-(\ell+w p_1)^2\big)
\widetilde\Phi_{J/\psi}\big(-(\ell+w p_2)^2\big)
\Tr\big[\gamma^\beta S(\ell+p_2)\gamma^\alpha S(\ell+p_1)\tilde\Gamma_{cJ} S(\ell)\big],
\nn
M^{\beta\alpha}_{\bigcirc\,b} &=& 
+ \int\frac{d^4\ell}{(2\pi)^4i}\int\limits_0^1\! d\tau\,
\widetilde\Phi^\prime_{cJ}(-z^+_\tau) 
\widetilde\Phi_{J/\psi}\big(-\ell^2\big)\,
\ell^\alpha\,\Tr\big[\gamma^\beta S(\ell+ w p_2)\tilde\Gamma_{cJ} S(\ell-w p_2)\big],
\nn
M^{\beta\alpha}_{\bigcirc\,c} &=& 
+\int\frac{d^4\ell}{(2\pi)^4i}\widetilde\Phi_{cJ}\big(-\ell^2\big)
\int\limits_0^1\! d\tau\,
\widetilde\Phi^\prime_{J/\psi}(-z^+_\tau)
\,\ell^\alpha\,\Tr\big[\gamma^\beta S(\ell+ w p_1)\tilde\Gamma_{cJ} S(\ell-w p_1)\big],
\nn[2ex]
z^+_\tau &=& (\ell+w q)^2\tau +\ell^2(1-\tau),
\nn
\tilde\Gamma_{c0} &=& I,\quad\tilde\Gamma_{c1}= \epsilon_\mu(p_1)\gamma^\mu\gamma_5,
\quad\tilde\Gamma_{c2} = 2\,\epsilon_{\mu\nu}(p_1)\ell^\mu \gamma^\nu,
\ena
where we have used the properties of transversality and symmetry of the polarization vectors:
$\epsilon^\ast_\alpha q^\alpha =0$, $\epsilon_{\mu\nu}=\epsilon_{\nu\mu}$, and $\epsilon_{\mu\nu} p_1^\mu =0$.

There is an additional diagram in the spin-2 case due to
the derivative in the Lagrangian. One has
\be
M^{\beta\alpha}_{\bigcirc\,d}  = 
- \int\frac{d^4\ell}{(2\pi)^4i} \widetilde\Phi_{cJ}\big(-(\ell+w q)^2\big)\widetilde\Phi_{J/\psi}(-\ell^2)\,
\epsilon_{\mu\nu}(p_1)\, g^{\nu\alpha}\Tr\big[\gamma^\beta S(\ell+ w p_2)\gamma^\mu S(\ell-w p_2)\big].
\en

The invariant matrix element describing the decay $h_c \to \eta_c + \gamma$ is written as
\bea
M_{h_c\to \eta_c + \gamma}  &=& 6\,g_{h_c}\,g_{\eta_c}\,e_q\,
\epsilon_\mu(p_1)\,\epsilon^\ast_{\gamma\,\alpha}(q)\,M^{\mu\alpha}\,,
\nn
M^{\mu\alpha} &=& 
M^{\mu\alpha}_{\triangle\,a} + M^{\mu\alpha}_{\bigcirc\,b} +  M^{\mu\alpha}_{\bigcirc\,c}  +  M^{\mu\alpha}_{\bigcirc\,d}, 
\nn
M^{\mu\alpha}_{\triangle\,a} &=& 
-\int\frac{d^4\ell}{(2\pi)^4i} \widetilde\Phi_{h_c}\Big(-(\ell+w p_1)^2\Big)
\widetilde\Phi_{\eta_c}\Big(-(\ell+w p_2)^2\Big),
\nn
&\times&
\Tr\Big[\gamma_5 S(\ell+p_2)\gamma^\alpha S(\ell+p_1) (2\ell^\mu)\gamma_5  S(\ell)\Big],
\nn
M^{\mu\alpha}_{\bigcirc\,b} &=& 
+ \int\frac{d^4\ell}{(2\pi)^4i}\int\limits_0^1 d\tau
\widetilde\Phi^\prime_{h_c}(-z^+_\tau)\, \ell^\alpha\,
\widetilde\Phi_{\eta_c}\Big(-\ell^2\Big)
\Tr\Big[\gamma_5 S(\ell+ w p_2)\gamma_5 (2\ell^\mu) S(\ell-w p_2)\Big],
\nn
M^{\mu\alpha}_{\bigcirc\,c} &=& 
+\int\frac{d^4\ell}{(2\pi)^4i}\widetilde\Phi_{h_c}\Big(-\ell^2\Big)
\int\limits_0^1 d\tau
\widetilde\Phi^\prime_{\eta_c}(-z^+_\tau)\,\ell^\alpha\,
\Tr\Big[\gamma_5 S(\ell+ w p_1)\gamma_5 (2\ell^\mu) S(\ell-w p_1)\Big],
\nn
M^{\mu\alpha}_{\bigcirc\,d} &=& 
- \int\frac{d^4\ell}{(2\pi)^4i} \widetilde\Phi_{h_c}\Big(-(\ell+w q)^2\Big)
\widetilde\Phi_{J/\psi}(-\ell^2)
\Tr\Big[\gamma_5 S(\ell+ w p_2)\gamma_5  g^{\mu\alpha} S(\ell-w p_2)\Big],
\nn[1.2ex]
z^+_\tau &=& (\ell+w q)^2\tau +\ell^2(1-\tau).
\nonumber
\ena

The first step is to check the gauge invariance before calculating the loop integration, i.e. to check
\be
M^{\beta\alpha}\,q_\alpha = 0.
\label{eq:gauge-con}
\en
It is done by using two identities:
\bea
S(\ell + p_2)\,\slash\!\!\! q\, S(\ell+p_1) &=& S(\ell+p_1) - S(\ell + p_2)\,,
\nn
\int\limits_0^1 \!\!d\tau \widetilde\Phi^\prime(-z_\tau)(\ell+w^2 q)^\alpha q_\alpha &=&
 \widetilde\Phi( -\ell^2 ) - \widetilde\Phi( -(\ell+w q)^2 ). 
 \nonumber
 \ena
 
The second step is to reduce the loop integrals to threefold integrals which are evaluated numerically.  The calculation of the matrix elements of the decays $\psi(2S)\to \chi_{cJ} + \gamma$ and $J/\psi\to\eta_c + \gamma$ is performed in a similar manner. 

 \section{Decay widths}
 \label{sec:Gam}

We assume that the charmonium and photon are on their mass shells. The  spin-1 polarization vectors $\epsilon^{(\lambda)}_\beta(p)$ and  $\epsilon^{(\lambda)}_\alpha(q)$ satisfy the conditions:
 \[
 \def\arraystretch{2.0}
  \begin{array}{ll}
 \epsilon^{(\lambda)}_\beta(p) \, p^\beta = 0, \qquad  \epsilon^{(\lambda)}_\alpha(q) \, q^\alpha = 0 & 
 \hspace{1cm}\text{transversality},
 \nn
 \sum\limits_{\lambda=0,\pm}\epsilon^{(\lambda)}_\beta(p)
\epsilon^{\dagger\,(\lambda)}_{\beta'} (p)
=-g_{\beta\beta'}+\frac{p_\beta\,p_{\beta'}}{m^2},  \qquad
\sum\limits_{\lambda=\pm}\epsilon^{(\lambda)}_\alpha(q)
\epsilon^{\dagger\,(\lambda)}_{\alpha'} (q)=-g_{\alpha\alpha'} & \hspace{1cm} \text{completeness},
\nn
\epsilon^{\dagger\,(\lambda)}_\mu \epsilon^{(\lambda')\,\mu} = -\delta_{\lambda \lambda'} &
\hspace{1cm} \text{orthonormality}.
\nonumber
\end{array}
\]
And the spin-2 polarization vector $\epsilon^{(\lambda)}_{\mu\nu}(p)$ satisfies
the conditions:
\[
\def\arraystretch{1.7}
\begin{array}{ll}
\displaystyle\epsilon^{(\lambda)}_{\mu\nu}(p) = \epsilon^{(\lambda)}_{\nu\mu}(p)&
\hspace*{1cm} {\rm  symmetry},
\\
\displaystyle\epsilon^{(\lambda)}_{\mu\nu}(p)\, p^\mu = 0 &
\hspace*{1cm} {\rm transversality},
\\
\displaystyle\epsilon^{(\lambda)}_{\mu\mu}(p) = 0 &
\hspace*{1cm} {\rm tracelessness},
\\
\displaystyle\sum\limits_{\lambda=0,\pm 1, \pm 2}\epsilon^{(\lambda)}_{\mu\nu}
\epsilon^{\dagger\,(\lambda)}_{\alpha\beta}
=\frac{1}{2}\left(S_{\mu\alpha}\,S_{\nu\beta}
                   +S_{\mu\beta}\,S_{\nu\alpha}\right)
  -\frac{1}{3}\,S_{\mu\nu}\,S_{\alpha\beta} &
\hspace*{1cm} {\rm completeness},
\\
\displaystyle\epsilon^{\dagger\,(\lambda)}_{\mu\nu} \epsilon^{(\lambda')\,\mu\nu}
=\delta_{\lambda \lambda'} &
\hspace*{1cm}  {\rm orthonormality},
\end{array}
\]
where
$ S_{\mu\nu} = -g_{\mu\nu}+\frac{p_\mu\,p_\nu}{m^2} $.

\begin{center}
{\bf Transition  \boldmath{$\chi_{c0}\to J/\psi+ \gamma$} }
\end{center}
\bea
M_{\chi_{c0}\to J/\psi+ \gamma} &=&
e\,\epsilon^\ast_{2\,\beta}(p_2)\,\epsilon^\ast_{\gamma\,\alpha}(q)\,
M_{\chi_{c0}}^{\beta\alpha}\,,
\nn
M_{\chi_{c0}}^{\beta\alpha} &=&
(m_1 A_1)\big( g^{\beta\alpha} - \frac{q^\beta p_2^\alpha}{p_2 q} \big) ,
\qquad p_2q =\frac{m_1^2-m_2^2}{2}\,,
\nn[2ex]
\Gamma(\chi_{c0}\to J/\psi+ \gamma) &=& \alpha\, \mathbf{|q|}\, A_1^2, \qquad
\mathbf{|q|} = \frac{m_1^2-m_2^2}{2m_1}.
\nonumber
\ena

\begin{center}
{\bf Transition \boldmath{$\chi_{c1}\to J/\psi+ \gamma$} }
\end{center}

There are two independent helicity amplitudes
$H_{\lambda_1;\lambda_2\lambda}$ which we denote by 
$H_{i}\,\,(i=L,T)$ according to the helicity of the final meson state $\Jpsi$, 
where $\lambda_2=0$ and $\lambda_2=\pm1$ stand for the longitudinal
and transverse helicities of $\Jpsi$. From parity one has 
$H_{+;0-}=-H_{-;0+}=H_{L}$ and $H_{0;++}=-H_{0;--}=H_{T}$.

The polarization vectors and momenta in the $\chi_{c1}-$rest frame
are defined as

\bea
\varepsilon_{1\,\mu}(\pm) &=& \tfrac{1}{\sqrt{2}}\, \Big(0;\pm 1,i,0\Big),\qquad
p_\mu = \Big(m_1; 0,0,0\Big),
\nn
\varepsilon_{1\,\mu}(0) &=& \Big(0;0,0,-1\Big), 
\nn[2ex]
\varepsilon^\dagger_{2\,\beta}(\pm) &=& \tfrac{1}{\sqrt{2}}\, 
\Big(0;\pm 1,-i,0\Big),\qquad
p_{2\,\beta} = \Big(E_2; 0,0,- \mathbf{|q|}\Big),
\nn
\varepsilon^\dagger_{2\,\beta}(0) &=& 
\tfrac{1}{m_2}\, \Big(\mathbf{| q|};0,0,-E_2\Big), \qquad  E_2 = \frac{m_1^2+m_2^2}{2m_1},
\nn[2ex]
\bar\varepsilon^\dagger_{\alpha}(\pm) &=& 
\tfrac{1}{\sqrt{2}}\, \Big(0;\mp 1,-i,0\Big),\qquad
q_\alpha = | \mathbf{q|}\Big(1;0,0,1\Big).
\label{eq:polS=1}
\ena

The $z$--direction is defined  by the momentum of $\Jpsi$.
The bars in the polarization four--vectors
$\bar{\varepsilon}_\alpha(\lambda)$ of the photon are a 
reminder that the photon helicities are defined relative to the negative 
$z$--direction.

The invariant matrix element is written in the form:
\bea
M_{\chi_{c1}\to J/\psi+ \gamma} &=&
e\,\epsilon_{1\,\mu}(p_1) \epsilon^\ast_{2\,\beta}(p_2)\,\epsilon^\ast_{\gamma\,\alpha}(q)\,
M_{\chi_{c1}}^{\mu\beta\alpha}\,,
\nn
M_{\chi_{c1}}^{\mu\beta\alpha} &=&
\big(\epsilon^{p_2q\mu\alpha}q^\beta\,D_E + \epsilon^{p_2q\beta\alpha}p_2^\mu\,D_M  \big) .
\label{eq:inv_amplC1}
\ena
It is convenient to present the decay width via the helicity amplitudes. One has

\bea
H_L &=& H_{+;0,-} = -  H_{-;0,+} =
\varepsilon_{1\,\mu}(+)\varepsilon^\dagger_{2\,\beta}(0)\bar\varepsilon^\dagger_{\alpha}(-)M_{\chi_{c1}}^{\mu\beta\alpha}
= i\frac{m_1^2}{m_2}\mathbf{|q|}^2 D_E,
\nn
H_T &=& H_{0;+,+} = -  H_{0;-,-} =
\varepsilon_{1\,\mu}(0)\varepsilon^\dagger_{2\,\beta}(+)\bar\varepsilon^\dagger_{\alpha}(+)M_{\chi_{c1}}^{\mu\beta\alpha}
= -i m_1\mathbf{|q|}^2 D_M .
\label{eq:helicity}
\ena

Then  the decay width is written as
\be
\Gamma(\chi_{c1}\to J/\psi+ \gamma) = \frac{\alpha}{3}\, \mathbf{|q|}\,
\Big( |H_L|^2 +  |H_T|^2 \Big).
\label{eq:width}
\en

\begin{center}
{\bf Transition \boldmath{$\chi_{c2}\to J/\psi+ \gamma$}}
\end{center}

There are three independent helicity amplitudes $H_{\lambda_1;\lambda_2\lambda}$
characterizing the decay $\chi_{c2}\to J/\psi+ \gamma$:
\bea
H_{+2;+1-1} &=& H_{-2;-1+1}=
\epsilon_{1\,\mu\nu}(+2) \epsilon^\ast_{2\,\beta}(+)\,\bar\epsilon^\ast_\alpha(-)\,
M_{\chi_{c2}}^{\mu\nu\beta\alpha}\,,
\nn
H_{+1;0-1} &=& H_{-1;0+1}=
\epsilon_{1\,\mu\nu}(+1) \epsilon^\ast_{2\,\beta}(0)\,\bar\epsilon^\ast_\alpha(-)\,
M_{\chi_{c2}}^{\mu\nu\beta\alpha}\,,
\nn
H_{0;+1+1} &=& H_{0;-1-1}=
\epsilon_{1\,\mu\nu}(0) \epsilon^\ast_{2\,\beta}(+)\,\bar\epsilon^\ast_\alpha(+)\,
M_{\chi_{c2}}^{\mu\nu\beta\alpha}\,.
\label{eq:helicity-S=2}
\ena

The polarization vectors of $\chi_{c2}$ are given by
\bea
\epsilon_{1\,\mu\nu}(\pm 2) &=&\epsilon_{2\,\mu}(\pm)\, \epsilon_{2\,\nu}(\pm),
\nn
\epsilon_{1\,\mu\nu}(\pm 1) &=&
\frac{1}{\sqrt{2}}\,\left( \epsilon_{2\,\mu}(\pm)\,\epsilon_{2\,\nu}(0)
                          +\epsilon_{2\,\mu}(0)\,\epsilon_{2\,\nu}(\pm)\right),
\nn
\epsilon_{1\,\mu\nu}(0) &=&
\frac{1}{\sqrt{6}}\,\left(\epsilon_{2\,\mu}(+)\,\epsilon_{2\,\nu}(-)
                       +\epsilon_{2\,\mu}(-)\,\epsilon_{2\,\nu}(+)\right)
+\sqrt{\frac{2}{3}}\,\epsilon_{2\,\mu}(0)\,\epsilon_{2\,\nu}(0)\,,
\label{eq:pol-S=2}
\ena
where $\epsilon_{2\,\mu}(r)$ are defined in Eq.~(\ref{eq:polS=1}).

The invariant matrix element $M_{\chi_{c2}}^{\mu\nu\beta\alpha}$ is calculated in the standard way
and represented in terms of the form factors. By using the gauge invariance the number of the form factors
is reduced to three. One has
\bea
M_{\chi_{c2}\to J/\psi+ \gamma} &=&
e\epsilon_{1\,\mu\nu}(p_1) \epsilon^\ast_{2\beta}(p_2)\epsilon^\ast_{\gamma\alpha}(q)
M_{\chi_{c2}}^{\mu\nu\beta\alpha},
\nn
M_{\chi_{c2}}^{\mu\nu\beta\alpha} &=&
F_1(p_2^\alpha q^\beta - p_2q g^{\alpha\beta} ) q^\mu q^\nu
+F_2( p_2^\alpha q^\nu - p_2q g^{\nu\alpha} )  g^{\mu\beta}\nn
&+& F_3(g^{\mu\alpha} q^\nu q^\beta - g^{\alpha\beta} q^\mu q^\nu ).
\label{eq:FF-S=2}
\ena

It is convenient to present the decay width via helicity amplitudes.
They are calculated by using the definitions given  in Eqs.~(\ref{eq:helicity-S=2})--(\ref{eq:FF-S=2}). One gets

\bea
\Gamma(\chi_{c2}\to J/\psi+ \gamma) &=& \frac{\alpha}{5}\frac{\mathbf{|q|}}{m_1^2}
\Big( |H_{+2;+1-1}|^2 + |H_{+1;0-1}|^2 + |H_{0;+1+1}|^2 \Big)\,,
\nn[2ex]
H_{+2;+1-1} &=& -m_1 \mathbf{|q|}\,F_2\,,
\nn
H_{+1;0-1} &=& -\frac{1}{\sqrt{2}}\frac{m_1}{m_2} \mathbf{|q|}\,\Big( E_2\,F_2 + \mathbf{|q|}\,F_3 \Big) \,,
\nn
H_{0;+1+1} &=& -\sqrt{\frac23} m_1\,\mathbf{|q|}\,\Big( \mathbf{|q|}^2 F_1 + \frac12\,F_2
+\frac{\mathbf{|q|}}{m_1}\,F_3 \Big) \,.
\label{eq:widthS2}  
  \ena

\begin{center}
{\bf Transition  \boldmath{$h_c\to \eta_c+ \gamma$} }
\end{center}
\bea
M_{h_c\to\eta_c + \gamma} &=&
e\,\epsilon_{1\,\mu}(p_1)\,\epsilon^\ast_{\gamma\,\alpha}(q)\,
M_{h_c}^{\mu\alpha}\,,
\nn
M_{h_c}^{\mu\alpha} &=&
W_1 \big( p_1^\alpha q^\mu - p_1q\, g^{\alpha\mu} \big) ,
\qquad p_1q =\frac{m_1^2-m_2^2}{2}\,,
\nn[2ex]
\Gamma(h_c\to \eta_c+ \gamma) &=& \frac{\alpha}{3}\, \mathbf{|q|^3}\, W_1^2, \qquad
\mathbf{|q|} = \frac{m_1^2-m_2^2}{2m_1}.
\ena

\section{Numerical results}
\label{sec:result}

The CCQM includes several free parameters: the constituent quark masses $m_q$, the hadron size parameters $\Lambda_H$, and the universal infrared cutoff parameter $\lambda$. In this paper, we apply a new strategy for fitting the model parameters. First, we have observed before that physical observables are rather insensitive to the concrete value of the cutoff parameter $\lambda$. In particular, in a recent study~\cite{Ganbold:2021nvj}, we checked this observation by gradually decreasing $\lambda$ and proved that the results did not change for any $\lambda < 0.181$~GeV up to the deconfinement limit. 
Therefore, in this study, we use again the value $\lambda = 0.181$~GeV obtained in our early fit~\cite{Branz:2009cd}. Additionally, by varying $\lambda$ by $\pm 10$~\% around the central value $0.181$~GeV we found that the results in this paper were not changed. We therefore ignore the error from this parameter in our analysis. The remaining free parameters include the charm quark mass $m_c$ and the size parameters $\Lambda_{cc}$ of the charmonium states. We observe that the charmonium radiative decay widths depend rather slowly on their size parameters $\Lambda_{cc}$ in the interval $2\sim 4$~GeV. Conversely, the dependence on the charm quark  mass $m_c$ running in the loop is strong. Therefore, we assume that the size parameter of a charmonium state is proportional to its mass, i.e. $\Lambda_{cc}=\rho M_{cc}$. 
We then determine $m_c$ and $\rho$ by fitting the calculated  branching fractions of the decays $J/\psi\to \eta_c\gamma$, $\chi_{c0}\to J/\psi\gamma$, and $\psi(2S)\to \chi_{c0}\gamma$ to experimental data. 

We use Bayesian inference for parameter estimation and Monte Carlo error propagation for predictions. We validate the Bayesian results using an independent, frequentist statistical approach. For the Bayesian analysis, we build a 16-dimensional parameter space that comprises two primary parameters of interest -- the charm quark mass $m_c$ and the hadron size parameter $\rho$ -- and 14 nuisance parameters $\{\alpha_j\}$ corresponding to the experimentally measured masses and widths of the charmonium states given in Table~\ref{tab:states}. The details of the fitting procedure and error propagation calculation are given in Appendix~\ref{sec:fit}. Here we present only the main results.

\subsection{Parameter constraints and model degeneracy}
\label{subsec:posterior}
The Bayesian analysis framework yields robust constraints on the model's two principal parameters. The median values and 68\% credible intervals, derived from the marginal posterior distributions, are determined to be:
\begin{align}
	m_c  &= {1.83}^{+0.01}_{-0.01}\,\textrm{GeV}, \nn
	\rho &= {0.61}^{+0.06}_{-0.05}. \label{eq:fit_result}
\end{align}
\begin{figure}[htbp]
	\centering
	\includegraphics[width=0.75\textwidth]{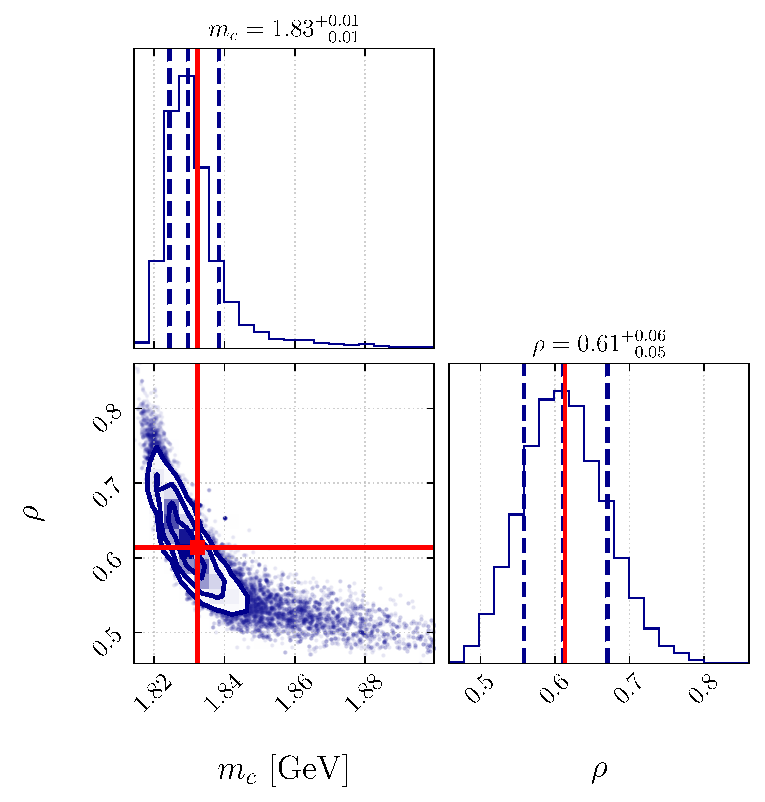} 
	\vspace*{-3mm}
	\caption{Posterior probability distributions for the charm quark mass $m_c$ and the size parameter $\rho$.}
	\label{fig:corner_plot}
\end{figure}
The full posterior probability distributions for these parameters are presented in Fig.~\ref{fig:corner_plot}. As illustrated, the diagonal panels show the one-dimensional marginalized posterior for each parameter, while the off-diagonal panel shows their two-dimensional joint probability. The contours on the joint distribution, which represent 0.5, 1.0, 1.5, and 2.0$\sigma$ credible regions, reveal a strong anti-correlation. This relationship is quantified by a Pearson coefficient of $r = -0.772$ and indicates a significant physical degeneracy within the model, where a smaller $m_c$ value can be compensated for by a larger $\rho$ to produce similar physical observables.

\subsection{Goodness-of-fit and model tension}
\label{subsec:gof}
\begin{figure}[htbp]
	\centering
	\includegraphics[width=0.8\textwidth]{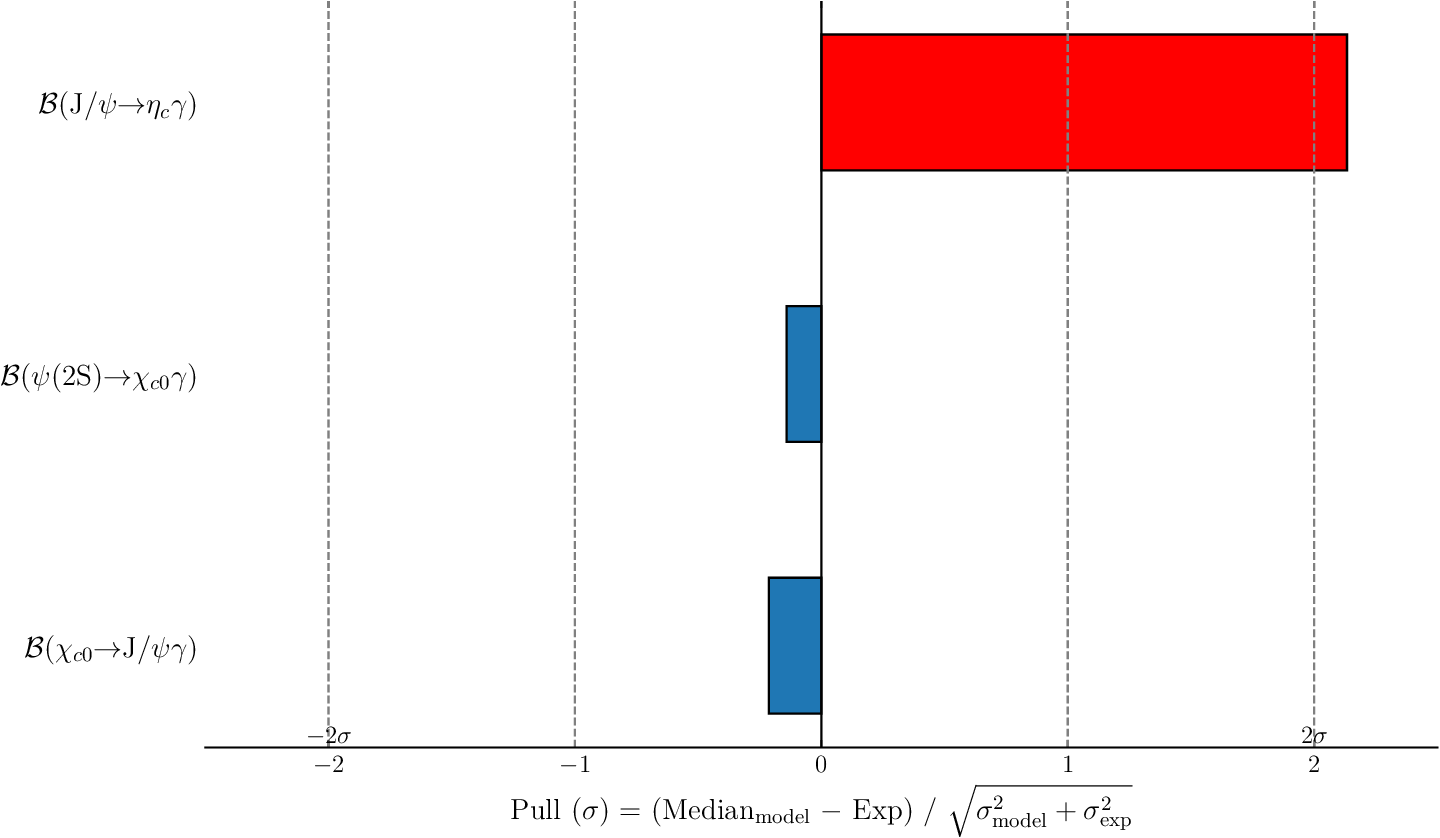}
	\vspace*{-3mm}
	\caption{Pull summary for the three fitted decay channels.}
	\label{fig:pull_summary}
\end{figure}

\begin{figure}[htbp]
	\begin{tabular}{lr}		
		\includegraphics[width=0.5\textwidth]{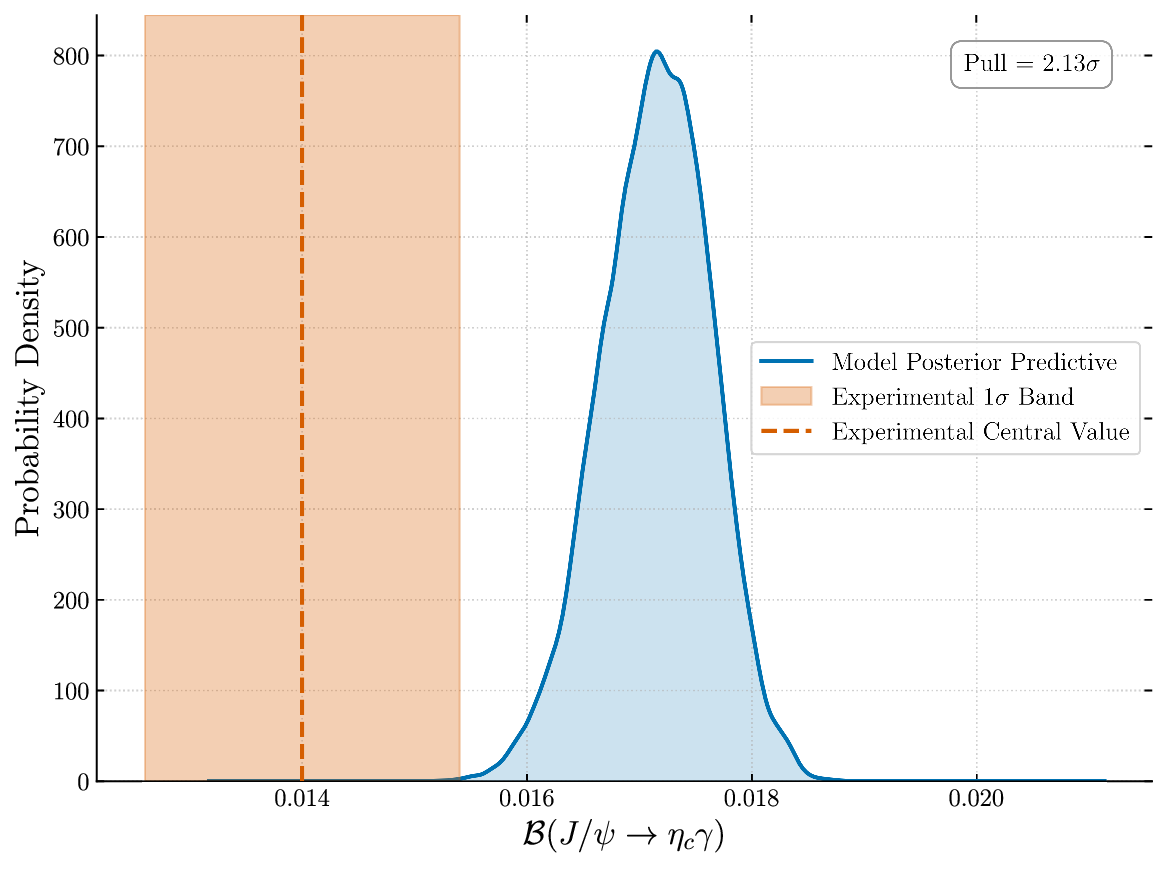}&\\
		\includegraphics[width=0.5\textwidth]{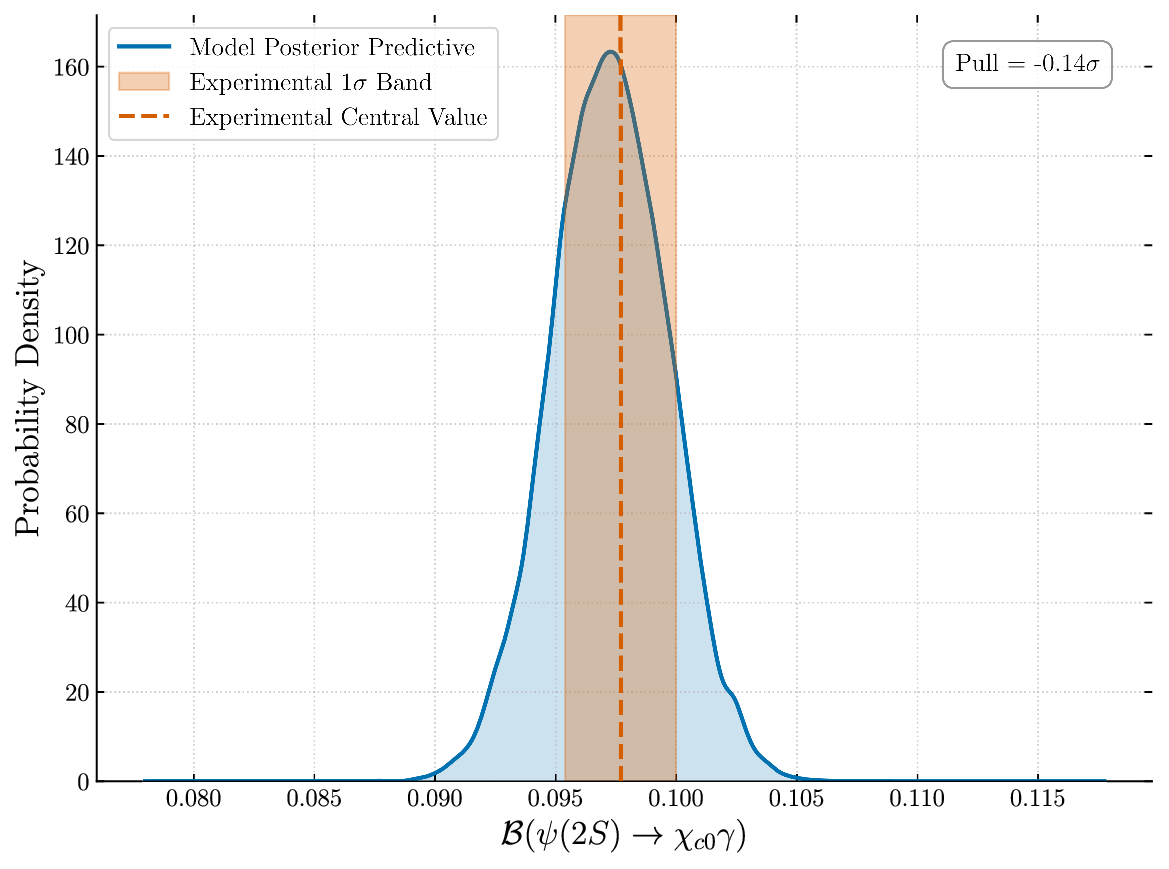}
		&
		\includegraphics[width=0.5\textwidth]{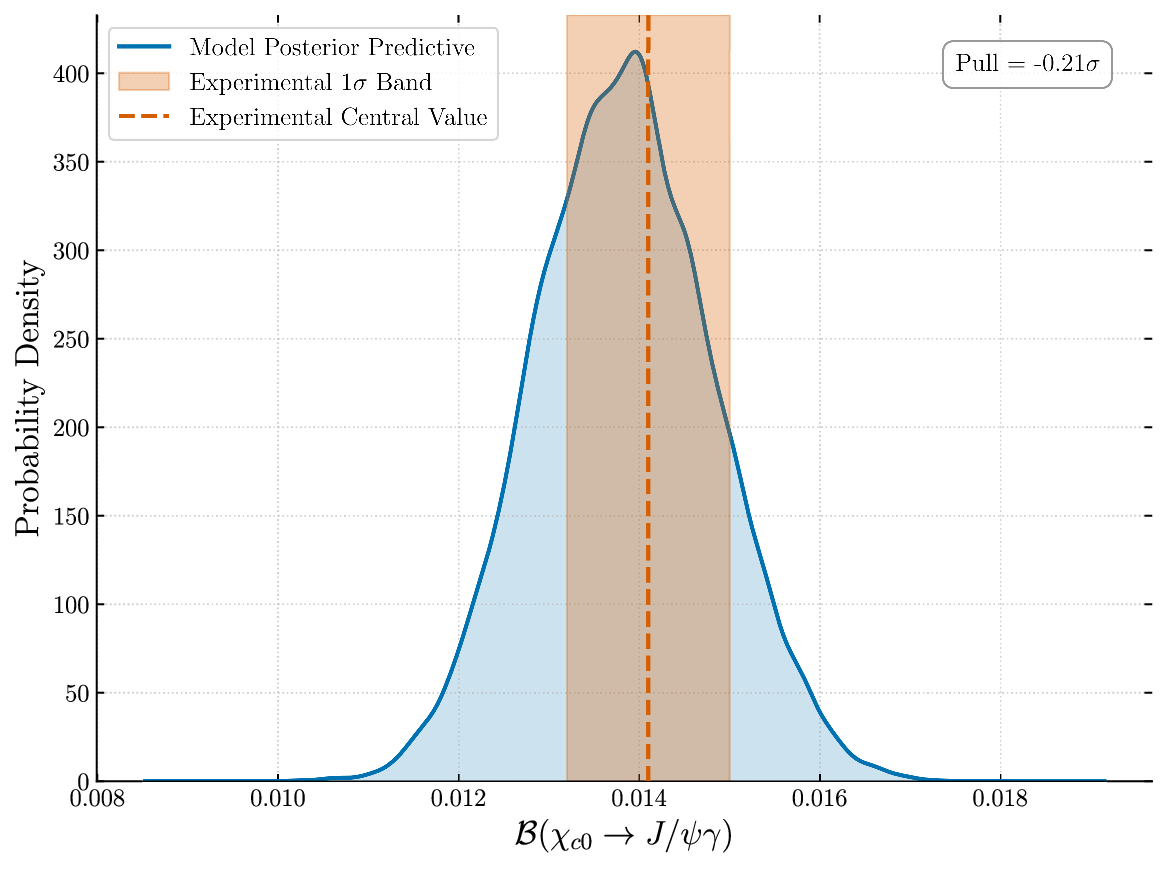}
	\end{tabular}
\vspace*{-3mm}
	\caption{Posterior predictive check for the fitted channels.}
	\label{fig:PPC}
\end{figure}

While the parameters are well-constrained, the model's overall performance exhibits a quantifiable tension with the data. The global goodness-of-fit, evaluated at the posterior's point of maximum likelihood, is characterized by:
\begin{equation}
	\chi^2_\text{min} / \text{d.o.f.} = 5.16 / 1.
\end{equation}
This value indicates that the model as a whole is a poor descriptor of the data. To diagnose the source of this tension, we examine the pull of each individual channel (Fig.~\ref{fig:pull_summary}). This decomposition makes it clear that the poor global fit is not a general failure of the model, but is instead localized to a single, large discrepancy in the $J/\psi \to \eta_c\gamma$ channel.
A detailed posterior predictive check (PPC) for the fitted channels, shown in Fig.~\ref{fig:PPC}, visually confirms this finding. In the plot, the model's posterior predictive distribution (blue histogram) for $\mathcal{B}(J/\psi \to \eta_c\gamma)$ is systematically offset from the experimental data (red line and 1$\sigma$ band), resulting in a large pull of $+2.13\sigma$. In contrast, the model demonstrates excellent agreement with experiments for the other two fitted channels. The model's results for $\psi(2S) \to \chi_{c0}\gamma$ and $\chi_{c0} \to J/\psi\gamma$ align remarkably well with the experimental data. Their pulls of $-0.14\sigma$ and $-0.21\sigma$, respectively, are negligible. This finding strongly suggests that the model's deficiency is localized to the description of the $J/\psi \to \eta_c\gamma$ transition, rather than being a global failure.

\subsection{Validation via frequentist cross-check}
\label{subsec:crosscheck}
\begin{figure}[H]
	\centering
	\includegraphics[width=\textwidth]{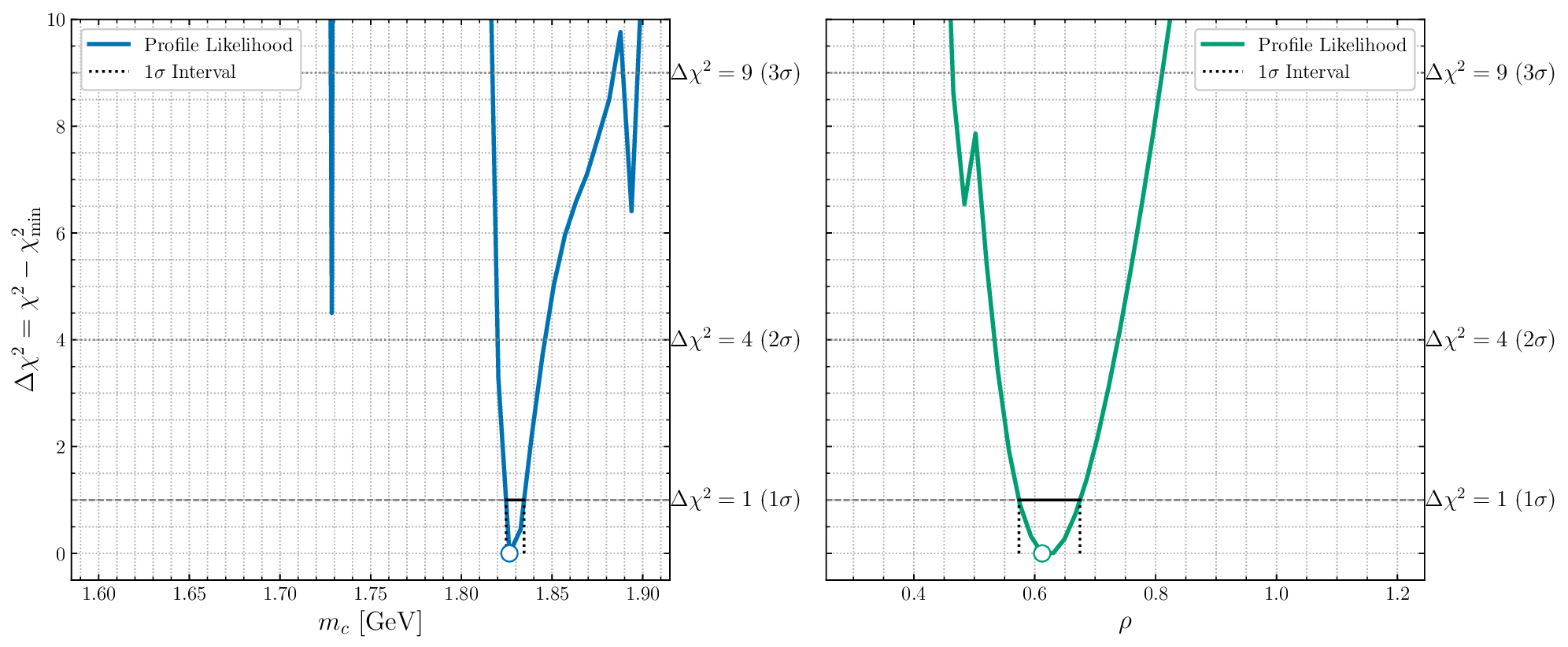}
	\caption{Frequentist profile likelihood scans for the $m_c$ and $\rho$ parameters.}
	\label{fig:profile_likelihood}
\end{figure}

To ensure the robustness of our findings, we performed an independent cross-check using the frequentist profile likelihood method. The resulting profile curves are shown in Fig.~\ref{fig:profile_likelihood}. The $1\sigma$ (68.3\%) confidence level (C.L.) intervals are defined by the standard criterion $\Delta\chi^2 = \chi^2 - \chi^2_\text{min} \le 1$. In the figure, the horizontal lines mark the 1$\sigma$, 2$\sigma$, and 3$\sigma$ confidence levels, corresponding to $\Delta\chi^2$ values of 1, 4, and 9, respectively.
This procedure yields the following frequentist estimates:
\begin{align}
	m_c  &\in [{1.82}, {1.84}]\,\textrm{GeV} & &(68.3\% \text{ C.L.}), \nn
	\rho &\in [{0.56}, {0.66}] & &(68.3\% \text{ C.L.}).
\end{align}
These confidence intervals are in excellent agreement with the Bayesian credible intervals reported in Eq.~(\ref{eq:fit_result}). This strong consistency between two distinct statistical paradigms confirms that our parameter estimates are robust and not an artifact of the chosen methodology.

\subsection{Theoretical prediction}
\label{subsec:predictions}
\begin{figure}[htpb]
	\centering
	\includegraphics[width=0.75\textwidth]{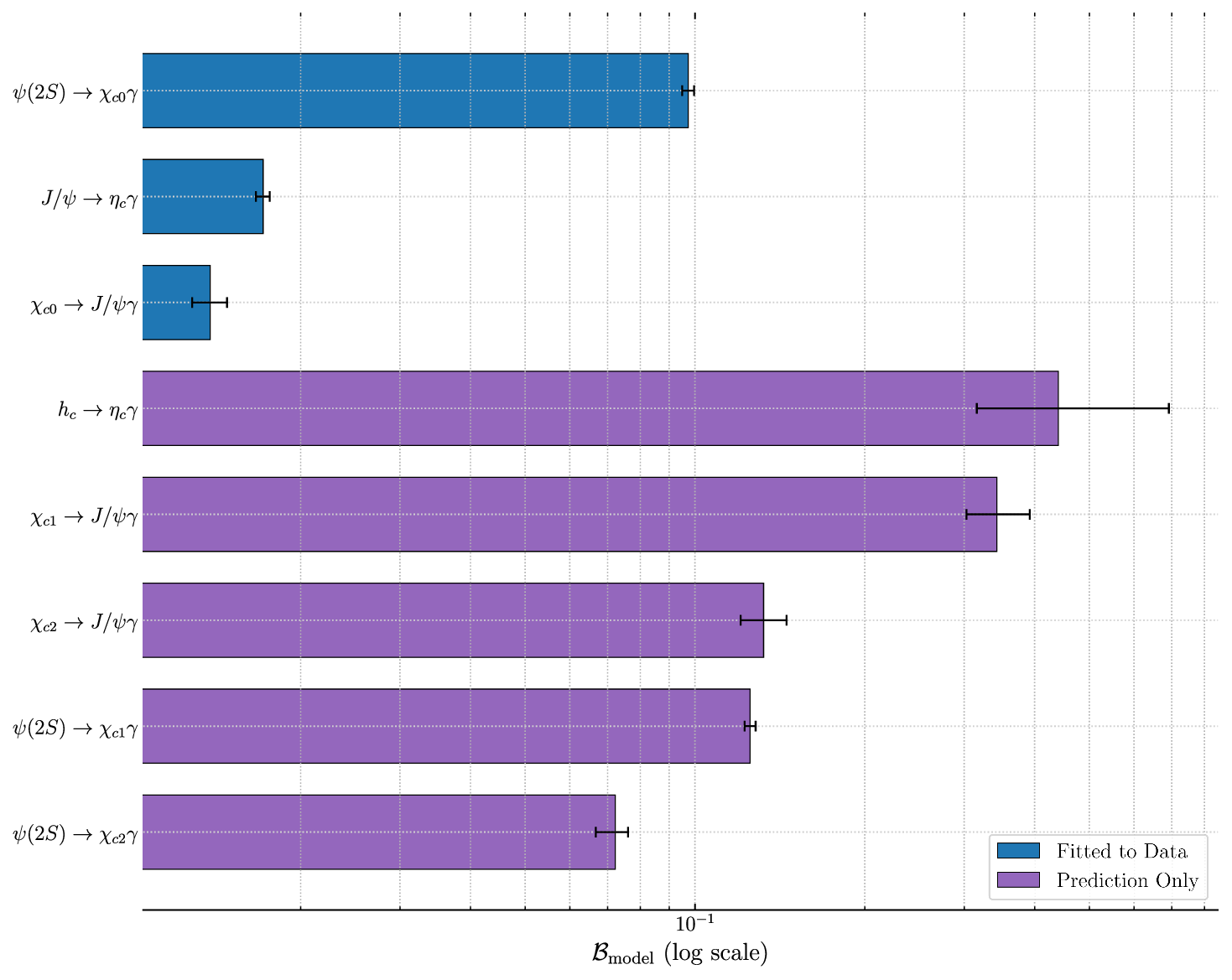}
	\vspace*{-3mm}
	\caption{Model predictions for branching fractions across all considered channels.}
	\label{fig:all_predictions}
\end{figure}

In Fig.~\ref{fig:all_predictions} we provide a comprehensive summary of the predicted branching fractions, visually distinguishing between fitted (blue) and prediction-only (purple) channels. A detailed breakdown of these predictions, along with a decomposition of their theoretical uncertainty, is summarized in Table~\ref{tab:predictions}.
\begin{table}[htbp]
	\centering	
	\begin{ruledtabular}	
	\begin{tabular}{cccccccc}		
		& \multicolumn{2}{c}{Prediction ($\mathcal{B}(10^{-2})$)} & & \multicolumn{3}{c}{Variance Contribution (\%)\textsuperscript{\emph{a}}}& \\
		\cmidrule(lr){2-3} \cmidrule(lr){5-7}
		Decay Channel & \multicolumn{1}{c}{Value} & \multicolumn{1}{c}{$\sigma_\text{tot}$} & & $m_c$ & $\rho$ & Sys. &Expt.~\cite{ParticleDataGroup:2024cfk}\\
		\midrule
		$h_c \to \eta_c\gamma$ & 39 & 32 & & 0.1 & 10.8 & \bfseries 89.1 & $60(4)$\\		
		$\chi_{c1} \to J/\psi\gamma$ & 32 & 9 & & 0.5 & \bfseries 79.3 & 20.2 & $34.3(1.3)$\\
		$\chi_{c2} \to J/\psi\gamma$ & 13 & 3 & & 1.8 & \bfseries 74.5 & 23.7 & $19.5(8)$\\
		$\psi(2S) \to \chi_{c1}\gamma$\textsuperscript{\emph{b}} & 12 & 11 & & \bfseries 92.8 & 5.9  & 1.3 & $9.75(27)$\\
		$\psi(2S) \to \chi_{c2}\gamma$ & 7 & 1 & & \bfseries 62.5 & 33.4 & 4.1 & $9.36(23)$\\
	\end{tabular}
	\end{ruledtabular}
\vspace*{-3mm}
	\caption{Model predictions and variance decomposition.}
	\label{tab:predictions}
	\begin{flushleft}
		\footnotesize
		\textsuperscript{\emph{a}} Percentage contribution to the total predictive variance, derived from the uncertainties in $m_c$, $\rho$, and all systematic (Sys.) nuisance parameters (Bold text indicates the largest contribution). \\
		\textsuperscript{\emph{b}} This channel requires high-precision numerical integration that proved unstable in the main analysis. This result was obtained via a dedicated recalculation using a robust error-handling framework that successfully managed isolated integration failures.
	\end{flushleft}
\end{table}

The variance decomposition (see~\ref{append:variance} for more details) in Table~\ref{tab:predictions} helps disentangle uncertainty sources. The channel $h_c \to \eta_c\gamma$ is dominated by systematic uncertainty (89.1\%). The class of $\psi(2S) \to \chi_{cJ}\gamma$ decays possess a high sensitivity to $m_c$ (92.8\% and 62.5\% for $J=1,2$) but a minimal dependence on $\rho$. In contrast, the decays $\chi_{c1,c2} \to J/\psi\gamma$ are more sensitive to $\rho$ (79.3\% and 74.5\% for $\chi_{c1}$, $\chi_{c2}$).  

Table~\ref{tab:predictions} shows good agreement between our predictions and experimental data. To be more specific, our predictions for the branching fractions $\mathcal{B}(h_c \to \eta_c\gamma)$, $\mathcal{B}(\chi_{c1} \to J/\psi\gamma)$, and $\mathcal{B}(\psi(2S) \to \chi_{c1}\gamma)$ agree with the corresponding experimental data within $1\sigma$. For the remaining decays $\chi_{c2} \to J/\psi\gamma$ and $\psi(2S) \to \chi_{c2}\gamma$, our results agree with data within $2\sigma$.
However, the theoretical uncertainties of our predictions in this study are large -- much larger than the estimated value of $\approx 10\%-20\%$  in our previous works (see, e.g.,~\cite{Tran:2024phq,Tran:2025fhb}). This difference is the consequence of (i) the full consideration of uncertainties in the mass and width of charmonium states (as can be seen in the case of $h_c \to \eta_c\gamma$), and (ii) the distict fitting strategy used in this study. In the traditional version of our model, each hadron $H$ is characterized by one size parameter  $\Lambda_H$, and all size parameters are assumed to be independent of each other. This means that the traditional description would require fitting 8 free parameters: the charm quark mass and 7 size parameters for 7 different charmonium states. In this study, we assumed that the size parameters of the charmonium states are proportional to their masses. This assumption reduces the number of free parameters from 8 to 2, which leads to improved predictive power but, as a trade-off, larger uncertainties in the theoretical predictions.

\section{Summary}
\label{sec:sum}
We present an analysis of the radiative decays of the charmonium states $h_c$, $J/\psi$, $\psi(2S)$, and $\chi_{cJ}$ ($J = \{0,1,2\}$) within the framework of our quark model (CCQM). We proved the gauge invariance of the amplitudes and obtained expressions for the decay width in terms of the helicity amplitudes. Assuming that the size parameters of charmonium states are proportional to their mass, $\Lambda_{cc} = \rho M_{cc}$, we performed a Bayesian fit for the charm quark mass $m_c$ and the coefficient of proportionality $\rho$. Three radiative channels $\psi(2S)\to \gamma\chi_{c0}$, $\chi_{c0}\to\gamma J/\psi$, and $J/\psi\to\gamma\eta_c$ were used to constrain the two parameters. The obtained values of $m_c$ and $\rho$ were then used to predict the branching fractions of the decays $h_c\to\gamma\eta_c$, $\chi_{c1,c2}\to\gamma J/\psi$, and $\psi(2S)\to\gamma\chi_{c1,c2}$. A Monte Carlo analysis was used to estimate the uncertainties of our theoretical predictions. Our predictions are in good agreement with available experimental data within uncertainty, which demonstrates the validity of the physical description of charmonium states in the CCQM.

\begin{acknowledgments}
This  research  is  funded  by  Vietnam National Foundation for Science and Technology Development (NAFOSTED) under grant number 103.01-2021.09. A.I.'s, M.A.I.'s, and Z.T.'s research has been funded by the Science Committee of the Ministry of Science and Higher Education of the Republic of Kazakhstan, Grant No. BR24992891.
\end{acknowledgments}

\appendix
\section{Fitting procedure and error estimation} 
\label{sec:fit}
The whole procedure is structured as a multi-stage analysis pipeline, designed to systematically constrain model parameters, evaluate model performance, and ensure the robustness of our predictions. This section details the methodology of each consecutive stage.

\subsection{Stage 1: Bayesian inference for parameter estimation}
\label{sec:stage1}

The objective of this stage is to perform a comprehensive exploration of the parameter space and determine the full joint posterior probability distribution for all model parameters. Our model operates in a 16-dimensional parameter space, denoted by the vector $\theta$. This space comprises two primary parameters of interest -- the charm quark mass $m_c$ and hadron size parameter $\rho$ -- and 14 nuisance parameters $\{\alpha_j\}$, corresponding to the measured masses and widths of the involved particles.

To navigate this high-dimensional space, we employ the \texttt{dynesty}~\cite{Speagle:2019ivv} package, a powerful Python implementation of the Nested Sampling algorithm. This method is particularly well-suited for this problem as it efficiently maps the posterior landscape, even in the presence of complex correlations, and simultaneously calculates the Bayesian evidence, which is crucial for model comparison.  The Bayesian framework requires two fundamental components: a likelihood function and a prior distribution.

\subsubsection{Likelihood function}

The likelihood function $\mathcal{L}(\theta | D)$ quantifies the probability of observing the experimental data $D$ given a specific set of model parameters $\theta$. In statistical analyses, it is often more convenient to work with the log-likelihood, which we define in terms of a total chi-squared function $\chi^2_{\textrm{total}}$:
\begin{equation}
	\ln \mathcal{L}(\theta | D) = -\frac{1}{2} \chi^2_{\textrm{total}}(\theta) + \textrm{const.}
\end{equation}
The $\chi^2_{\textrm{total}}$ is constructed from two distinct contributions: a goodness-of-fit term and a penalty term for the nuisance parameters, 
\begin{equation}
	\chi^2_{\textrm{total}}(\theta) = \chi^2_{\textrm{data}}(\theta) + \chi^2_{\textrm{penalty}}(\theta).
	\label{eq:chi2_total}
\end{equation}

The first term, $\chi^2_{\textrm{data}}$, is the conventional goodness-of-fit metric. It compares the theoretical branching fractions predicted by our model, $\mathcal{B}_{\textrm{th},i}(\theta)$, with the experimentally measured values, $\mathcal{B}_{\textrm{exp},i}$, across all $N$ decay channels used in the fit:
\begin{equation}
	\chi^2_{\textrm{data}}(\theta) = \sum_{i=1}^{N} \frac{\left( \mathcal{B}_{\textrm{th},i}(\theta) - \mathcal{B}_{\textrm{exp},i} \right)^2}{\sigma_{\textrm{exp},i}^2},
	\label{eq:chi2_data}
\end{equation}
where $\sigma_{\textrm{exp},i}$ is the uncertainty of the experimental data for the $i$-th channel.
The experimental values for $\mathcal{B}_{\textrm{exp},i}$ and $\sigma_{\textrm{exp},i}$ are taken from the PDG~\cite{ParticleDataGroup:2024cfk} and listed in Table~\ref{tab:exp_inputs}.
\begin{table}[H]
	\centering	
	\begin{tabular}{ccc}
		\hline\hline
		Decay Channel & {$\cal{B}_\textrm{exp}$} & {$\sigma_\textrm{exp}$} \\
		\hline
		$J/\psi \to \eta_c\gamma$        & 0.0140 & 0.0014 \\
		$\chi_{c0} \to J/\psi\gamma$     & 0.0141 & 0.0009 \\
		$\psi(2S) \to \chi_{c0}\gamma$   & 0.0977 & 0.0023 \\
		\hline\hline
	\end{tabular}
	\caption{Experimental inputs for the fitted decay channels.}
	\label{tab:exp_inputs}
\end{table}

The second term, $\chi^2_{\textrm{penalty}}$, constrains the 14 nuisance parameters \{$\alpha_j$\} to their established values from the PDG~\cite{ParticleDataGroup:2024cfk}. It takes the form of a sum of squared pulls, effectively incorporating external experimental information as Gaussian constraints within the likelihood itself:
\begin{equation}
	\chi^2_{\textrm{penalty}}(\theta) = \sum_{j=1}^{14} \frac{\left( \alpha_j - \bar{\alpha}_j \right)^2}{\sigma_{\alpha_j}^2},
	\label{eq:chi2_penalty}
\end{equation}
where $\bar{\alpha}_j$ and $\sigma_{\alpha_j}$ are the central value and uncertainty for the $j$-th nuisance parameter, respectively (see Table~\ref{tab:states}).

\subsubsection{Prior distributions}

A key aspect of our Bayesian approach is the choice of prior probability distributions $p(\theta)$ which represent our state of knowledge about the parameters before considering the data. For all 16 parameters in the model, we assigned uninformative, uniform (or ``flat") priors over a broad but physically plausible range $[\theta_{k, \min}, \theta_{k, \max}]$:
\begin{equation}
	p(\theta_k) = 
	\begin{cases} 
		\frac{1}{\theta_{k, \max} - \theta_{k, \min}} & \textrm{if } \theta_k \in [\theta_{k, \min}, \theta_{k, \max}], \\
		0 & \textrm{otherwise}.
	\end{cases}
	\label{eq:uniform_prior}
\end{equation}
The motivation for this choice is to adopt a maximally agnostic stance, allowing the data to dominate the inference process. Therefore, the final shape of the posterior distribution is dictated almost entirely by the likelihood function. By enforcing the strong constraints on the nuisance parameters via the $\chi^2_{\textrm{penalty}}$ term rather than through narrow priors, we improve the sampler's efficiency in exploring the valid regions of high likelihood.

\subsubsection{Outcome}

The successful execution of the Nested Sampling run yields a set of weighted samples which provides a discrete representation of the full 16-dimensional posterior probability distribution, $P(\theta|D) \propto \mathcal{L}(\theta|D) p(\theta)$. These results, which encapsulate all information about parameter correlations, uncertainties, and best-fit values, are serialized and saved to a file (\texttt{dynesty\_results.pkl}). This data object serves as the direct and complete input for Stage 2 of the analysis pipeline, where we perform post-processing, error propagation, and goodness-of-fit checks.

\subsection{Stage 2: Post-processing and error propagation}
\label{sec:stage2}

The output of the Nested Sampling algorithm is a discrete, weighted representation of the joint posterior probability distribution. The purpose of Stage 2 is to translate these raw statistical results into meaningful physical quantities, including central values and asymmetric uncertainties for the model parameters, and a comprehensive uncertainty budget for the model's predictions.

\subsubsection{Parameter uncertainty estimation}

The properties of each parameter are extracted from its one-dimensional marginal posterior distribution, $P(\theta_k|D)$, which is obtained by integrating the full joint posterior over all other parameters. This non-parametric approach correctly captures any asymmetries or non-Gaussian features present in the distributions.

For each parameter $\theta_k$, the central value, or bestfit point, is taken to be the median ($\hat{\theta}_k$) of its marginal posterior. The median, corresponding to the 50th percentile, is chosen for its robustness against skewed distributions, providing a more stable estimator of location compared to the mean. It is formally defined by the value $\hat{\theta}_k$ that satisfies
\begin{equation}
	\int_{-\infty}^{\hat{\theta}_k} P(\theta_k'|D) \,d\theta_k' = 0.5\,.
\end{equation}
The uncertainty is reported as the 68.3\% credible interval, corresponding to a $1\sigma$ range for a Gaussian distribution. This interval is defined by the 15.87th and 84.13th percentiles of the distribution, denoted $\theta_k^{(0.1587)}$ and $\theta_k^{(0.8413)}$, respectively. The asymmetric errors are then calculated as
\begin{align}
	\textrm{Upper Error: } +\sigma_{k}^{\textrm{up}} &= \theta_k^{(0.8413)} - \hat{\theta}_k, \label{eq:error_up} \\
	\textrm{Lower Error: } -\sigma_{k}^{\textrm{down}} &= \hat{\theta}_k - \theta_k^{(0.1587)}. \label{eq:error_down}
\end{align}
The final result for each parameter is reported as $\hat{\theta}_k \substack{+\sigma_{k}^{\textrm{up}} \\ -\sigma_{k}^{\textrm{down}}}$. This method provides an honest representation of the parameter's uncertainty by faithfully reflecting the true shape of its posterior distribution.

\subsubsection{Monte Carlo error propagation for predictions}
\label{sec:mc_propagation}

To determine the theoretical uncertainty of a predicted quantity $y$ (e.g., a branching fraction not used in the fit), which is a function of the model parameters, $y = f(\theta)$, we must propagate the uncertainties from the full 16-dimensional parameter space. A Monte Carlo approach is ideally suited for this task as it naturally accounts for all parameter correlations. The procedure is executed as a formal algorithm:

\begin{enumerate}
	\item \textbf{Sampling from the posterior:} We draw a large number of random samples, $\{\theta^{(i)}\}_{i=1}^N$ with $N=20000$, from the joint posterior distribution $P(\theta|D)$ stored in \texttt{dynesty\_results.pkl}. Each sample $\theta^{(i)}$ is a 16-dimensional vector representing a complete, physically plausible parameter set.
	
	\item \textbf{Iterative calculation:} For each sample vector $\theta^{(i)}$, we execute the physics model $f$ to calculate the corresponding predicted value: $y^{(i)} = f(\theta^{(i)})$.
	
	\item \textbf{Generating the prediction distribution:} This process yields a set of $N$ predicted values, $\{y^{(i)}\}_{i=1}^N$. This set forms an empirical representation of the posterior predictive distribution for the quantity $y$.
	
	\item \textbf{Extracting the final result:} The central value of the prediction $\hat{y}$ is calculated using the bestfit (median) parameter vector $\hat{\theta}$:
	\begin{equation}
		\hat{y} = f(\hat{\theta}).
	\end{equation}
	The total model uncertainty $\sigma_y$ for this prediction  is taken as the standard deviation of the distribution of predicted values:
	\begin{equation}
		\sigma_y = \textrm{std}\left( \{y^{(i)}\}_{i=1}^N \right).
	\end{equation}
	This propagated error comprehensively includes all sources of uncertainty and the complex, nonlinear correlations between them.
\end{enumerate}

\subsubsection{Decomposition of the uncertainty budget}
\label{append:variance}

To understand the relative impact of different sources of uncertainty of our predictions, we decompose the total uncertainty into its primary constituents using a parameter-freezing technique. To isolate the uncertainty contribution from a single parameter, e.g. $m_c$, we modify the Monte Carlo procedure as follows. In each of the $N$ sampled vectors $\theta^{(i)} = (m_c^{(i)}, \rho^{(i)}, \{\alpha_j^{(i)}\})$, only the $m_c$ component is allowed to vary according to its posterior draws, while all other 15 parameters are held fixed at their best-fit (median) values ($\hat{\rho}, \{\hat{\alpha}_j\}$). The uncertainty contribution from $m_c$ alone, denoted $\sigma_{m_c}$, is the standard deviation of the resulting distribution of predictions:
\begin{equation}
	\sigma_{m_c} = \textrm{std} \left( \left\{ f(m_c^{(i)}, \hat{\rho}, \{\hat{\alpha}_j\}) \right\}_{i=1}^N \right).
\end{equation}
This process is repeated for the parameter $\rho$ (yielding $\sigma_{\rho}$) and for the block of 14 experimental nuisance parameters (yielding the systematic uncertainty $\sigma_{\textrm{syst}}$). Because the posterior distributions are not perfectly Gaussian and the model may be nonlinear, the variances do not sum exactly. However, they provide an excellent approximation of the total uncertainty budget, which can be checked for closure:
\begin{equation}
	\sigma_{y}^2 \approx \sigma_{m_c}^2 + \sigma_{\rho}^2 + \sigma_{\textrm{syst}}^2.
\end{equation}
This decomposition allows us to quantify the percentage contribution of each source, thereby identifying which parameters are the dominant drivers of uncertainty in our theoretical predictions.

\subsection{Stage 3: Frequentist cross-check with profile likelihood}
\label{sec:stage3}

While the Bayesian analysis in Stage 1 was used for fitting the model parameters, we perform a validation using an independent, frequentist statistical approach. The objective is to corroborate the Bayesian credible intervals with frequentist confidence intervals calculated via the profile likelihood method. A strong agreement between the results from these two distinct paradigms significantly enhances the reliability of our conclusions. This stage relies on numerical optimization routines, for which we employ the \texttt{scipy.optimize} library~\cite{Virtanen:2019joe}.

\subsubsection{The profile likelihood method}

The profile likelihood method provides a rigorous framework for constructing confidence intervals for a single parameter of interest (POI) while systematically accounting for the uncertainties associated with all other nuisance parameters. Let the full 16-dimensional parameter vector be denoted $\theta$. To construct an interval for a POI, e.g. $\Psi = m_c$, we partition the vector as $\theta = (\Psi, \lambda)$, where $\lambda$ represents the remaining 15 nuisance parameters (including $\rho$ and the 14 experimental nuisances).

First, the global minimum of the $\chi^2_{\textrm{total}}$ function is found by minimizing with respect to all parameters simultaneously. The bestfit parameter vector $\hat{\theta}$ and the minimum $\chi^2$ value are given by:
\begin{align}
	\hat{\theta} &= \arg\underset{\theta}{\min} \, \chi^2_{\textrm{total}}(\theta), \\
	\chi^2_{\textrm{global,min}} &= \chi^2_{\textrm{total}}(\hat{\theta}).
\end{align}
Next, the profile $\chi^2$ function, $\chi^2_{\textrm{prof}}(\Psi)$, is constructed. For each fixed value of the POI $\Psi$ on a predefined grid, the function $\chi^2_{\textrm{total}}(\Psi, \lambda)$ is minimized with respect to all nuisance parameters $\lambda$. This procedure, known as profiling, defines the conditional minimum:
\begin{equation}
	\chi^2_{\textrm{prof}}(\Psi) = \min_{\lambda} \chi^2_{\textrm{total}}(\Psi, \lambda).
\end{equation}
By tracing these conditional minima across the grid of $\Psi$ values, we obtain the profile $\chi^2$ curve. The quantity of interest is the delta chi-squared, $\Delta\chi^2(\Psi)$, which measures the increase in $\chi^2$ from its global minimum as the POI is varied:
\begin{equation}
	\Delta\chi^2(\Psi) = \chi^2_{\textrm{prof}}(\Psi) - \chi^2_{\textrm{global,min}}.
	\label{eq:delta_chi2}
\end{equation}
This entire process is performed independently for each POI, such as $m_c$ and $\rho$.

\subsubsection{Interval estimation and validation}

According to Wilks' theorem~\cite{Wilks:1938dza}, for a single parameter of interest, the $\Delta\chi^2$ statistic asymptotically follows a $\chi^2$ distribution with one degree of freedom ($k=1$). This provides a direct mapping between values of $\Delta\chi^2$ and standard confidence levels (C.L.) in the frequentist framework. The $N\sigma$ confidence interval is defined by the range of POI values for which the $\Delta\chi^2$ curve remains below $N^2$. Specifically:
\begin{itemize}
	\item The 1$\sigma$ (68.3\% C.L.) interval is found by solving $\Delta\chi^2(\Psi) = 1.0$.
	\item The 2$\sigma$ (95.4\% C.L.) interval is found by solving $\Delta\chi^2(\Psi) = 4.0$.
\end{itemize}
The final validation step consists of a direct comparison between these frequentist confidence intervals and the Bayesian 68.3\% credible intervals derived in Stage 2. A close agreement between these independently derived uncertainty bounds confirms that our results are not an artifact of the chosen statistical methodology and that our uncertainty quantification is robust.

\ed
\begin{thebibliography}{99}

\bibitem{Brambilla:2010cs}
  N.~Brambilla, S.~Eidelman, B.~K.~Heltsley, R.~Vogt, G.~T.~Bodwin, E.~Eichten, A.~D.~Frawley,
  A.~B.~Meyer, R.~E.~Mitchell, V.~Papadimitriou \textit{et al.}
Eur. Phys. J. C \textbf{71}, 1534 (2011)
[arXiv:1010.5827 [hep-ph]].


\bibitem{Gaiser:1985ix}
  J.~Gaiser, E.~D.~Bloom, F.~Bulos, G.~Godfrey, C.~M.~Kiesling,
  W.~S.~Lockman, M.~Oreglia, D.~L.~Scharre, C.~Edwards, R.~Partridge \textit{et al.}
Phys. Rev. D \textbf{34}, 711 (1986).


\bibitem{FermilabE835:2002pkx}
S.~Bagnasco \textit{et al.} (Fermilab E835 Collaboration),
Phys. Lett. B \textbf{533}, 237 (2002).

\bibitem{CLEO:2004cbu}
S.~B.~Athar \textit{et al.} (CLEO Collaboration),
Phys. Rev. D \textbf{70}, 112002 (2004)
[arXiv:hep-ex/0408133 [hep-ex]].

\bibitem{CLEO:2005efp}
N.~E.~Adam \textit{et al.} (CLEO Collaboration),
Phys. Rev. Lett. \textbf{94}, 232002 (2005)
[arXiv:hep-ex/0503028 [hep-ex]].



\bibitem{CLEO:2008kwj}
H.~Mendez \textit{et al.} (CLEO Collaboration),
Phys. Rev. D \textbf{78}, 011102 (2008)
[arXiv:0804.4432 [hep-ex]].

\bibitem{CLEO:2005vqq}
J.~L.~Rosner \textit{et al.} (CLEO Collaboration),
Phys. Rev. Lett. \textbf{95}, 102003 (2005)
[arXiv:hep-ex/0505073 [hep-ex]].

\bibitem{CLEO:2008ero}
S.~Dobbs \textit{et al.} (CLEO Collaboration),
Phys. Rev. Lett. \textbf{101}, 182003 (2008)
[arXiv:0805.4599 [hep-ex]].

\bibitem{CLEO:2008pln}
R.~E.~Mitchell \textit{et al.} (CLEO Collaboration),
Phys. Rev. Lett. \textbf{102}, 011801 (2009)
[erratum: Phys. Rev. Lett. \textbf{106}, 159903 (2011)]
[arXiv:0805.0252 [hep-ex]].

\bibitem{BES:2005bmx}
M.~Ablikim \textit{et al.} (BES Collaboration),
Phys. Rev. D \textbf{71}, 092002 (2005)
[arXiv:hep-ex/0502031 [hep-ex]].


\bibitem{BESIII:2012urf}
M.~Ablikim \textit{et al.} (BESIII Collaboration),
Phys. Rev. D \textbf{86}, 092009 (2012)
[arXiv:1209.4963 [hep-ex]].

\bibitem{BESIII:2017gcu}
M.~Ablikim \textit{et al.} (BESIII Collaboration),
Phys. Rev. D \textbf{96},  032001 (2017)
[arXiv:1703.00077 [hep-ex]].

\bibitem{BESIII:2010gid}
M.~Ablikim \textit{et al.} (BESIII Collaboration),
Phys. Rev. Lett. \textbf{104}, 132002 (2010)
[arXiv:1002.0501 [hep-ex]].

\bibitem{BESIII:2022tfo}
M.~Ablikim \textit{et al.} (BESIII Collaboration),
Phys. Rev. D \textbf{106},  072007 (2022)
[arXiv:2204.09413 [hep-ex]].

\bibitem{ParticleDataGroup:2024cfk}
  S.~Navas \textit{et al.} [Particle Data Group],
Phys. Rev. D \textbf{110},  030001 (2024)


\bibitem{Barnes:2005pb}
T.~Barnes, S.~Godfrey, and E.~S.~Swanson,
Phys. Rev. D \textbf{72}, 054026 (2005)
[arXiv:hep-ph/0505002 [hep-ph]].

\bibitem{Eichten:2007qx}
E.~Eichten, S.~Godfrey, H.~Mahlke, and J.~L.~Rosner,
Rev. Mod. Phys. \textbf{80}, 1161 (2008)
[arXiv:hep-ph/0701208 [hep-ph]].

\bibitem{Voloshin:2007dx}
M.~B.~Voloshin,
Prog. Part. Nucl. Phys. \textbf{61}, 455 (2008)
[arXiv:0711.4556 [hep-ph]].

\bibitem{Li:2009zu}
B.~Q.~Li and K.~T.~Chao,
Phys. Rev. D \textbf{79}, 094004 (2009)
[arXiv:0903.5506 [hep-ph]].

\bibitem{Cao:2012du}
L.~Cao, Y.~C.~Yang, and H.~Chen,
Few Body Syst. \textbf{53}, 327-342 (2012)
[arXiv:1206.3008 [hep-ph]].

\bibitem{Deng:2016stx}
W.~J.~Deng, H.~Liu, L.~C.~Gui, and X.~H.~Zhong,
Phys. Rev. D \textbf{95},  034026 (2017)
[arXiv:1608.00287 [hep-ph]].

\bibitem{Dudek:2006ej}
J.~J.~Dudek, R.~G.~Edwards, and D.~G.~Richards,
Phys. Rev. D \textbf{73}, 074507 (2006)
[arXiv:hep-ph/0601137 [hep-ph]].

\bibitem{Dudek:2009kk}
J.~J.~Dudek, R.~Edwards, and C.~E.~Thomas,
Phys. Rev. D \textbf{79}, 094504 (2009)
[arXiv:0902.2241 [hep-ph]].

\bibitem{Chen:2011kpa}
Y.~Chen, D.~C.~Du, B.~Z.~Guo, N.~Li, C.~Liu, H.~Liu, Y.~B.~Liu, J.~P.~Ma, X.~F.~Meng, Z.~Y.~Niu \textit{et al.}
Phys. Rev. D \textbf{84}, 034503 (2011)
[arXiv:1104.2655 [hep-lat]].

\bibitem{Becirevic:2025ocx}
D.~Be{\v{c}}irevi{\'c}, R.~Di Palma, R.~Frezzotti, G.~Gagliardi, V.~Lubicz, F.~Sanfilippo, and N.~Tantalo,
Phys. Rev. D \textbf{112},  034505 (2025)
[arXiv:2504.16807 [hep-lat]].

\bibitem{Khodjamirian:1979fa}
A.~Y.~Khodjamirian,
Phys. Lett. B \textbf{90}, 460 (1980)

\bibitem{Beilin:1984pf}
V.~A.~Beilin and A.~V.~Radyushkin,
Nucl. Phys. B \textbf{260}, 61 (1985)

\bibitem{DeFazio:2008xq}
F.~De Fazio,
Phys. Rev. D \textbf{79}, 054015 (2009)
[erratum: Phys. Rev. D \textbf{83}, 099901 (2011)]
[arXiv:0812.0716 [hep-ph]].

\bibitem{Wang:2012ph}
Z.~G.~Wang,
Int. J. Theor. Phys. \textbf{53}, 1022 (2014)
[arXiv:1206.5685 [hep-ph]].

\bibitem{Brambilla:2005zw}
N.~Brambilla, Y.~Jia, and A.~Vairo,
Phys. Rev. D \textbf{73}, 054005 (2006)
[arXiv:hep-ph/0512369 [hep-ph]].

\bibitem{Brambilla:2012be}
N.~Brambilla, P.~Pietrulewicz, and A.~Vairo,
Phys. Rev. D \textbf{85}, 094005 (2012)
[arXiv:1203.3020 [hep-ph]].

\bibitem{Pineda:2013lta}
A.~Pineda and J.~Segovia,
Phys. Rev. D \textbf{87}, 074024 (2013)
[arXiv:1302.3528 [hep-ph]].

\bibitem{Ebert:2002pp}
D.~Ebert, R.~N.~Faustov, and V.~O.~Galkin,
Phys. Rev. D \textbf{67}, 014027 (2003)
[arXiv:hep-ph/0210381 [hep-ph]].

\bibitem{Bruschini:2019vff}
R.~Bruschini and P.~Gonz\'alez,
Phys. Rev. D \textbf{101}, no.1, 014027 (2020)
[arXiv:1910.06773 [hep-ph]].

\bibitem{Ke:2013zs}
H.~W.~Ke, X.~Q.~Li, and Y.~L.~Shi,
Phys. Rev. D \textbf{87},  054022 (2013)
[arXiv:1301.4014 [hep-ph]].

\bibitem{Guo:2014zva}
P.~Guo, T.~Y\'epez-Mart\'\i{}nez, and A.~P.~Szczepaniak,
Phys. Rev. D \textbf{89},  116005 (2014)
[arXiv:1402.5863 [hep-ph]].


\bibitem{Ganbold:2021nvj}
G.~Ganbold, T.~Gutsche, M.~A.~Ivanov, and V.~E.~Lyubovitskij,
Phys. Rev. D \textbf{104},  094048 (2021)
[arXiv:2107.08774 [hep-ph]].


\bibitem{Mandelstam:1962mi}
S.~Mandelstam,
Annals Phys. \textbf{19}, 1 (1962)

\bibitem{Terning:1991yt}
J.~Terning,
Phys. Rev. D \textbf{44},  887 (1991)

\bibitem{Ivanov:1996pz}
M.~A.~Ivanov, M.~P.~Locher, and V.~E.~Lyubovitskij,
Few Body Syst. \textbf{21}, 131 (1996)
[arXiv:hep-ph/9602372 [hep-ph]].

\bibitem{Ivanov:1998wj}
M.~A.~Ivanov, J.~G.~K\"orner, and V.~E.~Lyubovitskij,
Phys. Lett. B \textbf{448}, 143 (1999)
[arXiv:hep-ph/9811370 [hep-ph]].

\bibitem{Ivanov:1999bk}
M.~A.~Ivanov, J.~G.~K\"orner, V.~E.~Lyubovitskij, and A.~G.~Rusetsky,
Phys. Rev. D \textbf{60}, 094002 (1999)
[arXiv:hep-ph/9904421 [hep-ph]].

\bibitem{Branz:2009cd}
  T.~Branz, A.~Faessler, T.~Gutsche, M.~A.~Ivanov,
  J.~G.~K\"orner, and V.~E.~Lyubovitskij,
Phys. Rev. D \textbf{81}, 034010 (2010)
[arXiv:0912.3710 [hep-ph]].

\bibitem{Branz:2010pq}
T.~Branz, A.~Faessler, T.~Gutsche, M.~A.~Ivanov, J.~G.~K\"orner, V.~E.~Lyubovitskij, and B.~Oexl,
Phys. Rev. D \textbf{81}, 114036 (2010)
[arXiv:1005.1850 [hep-ph]].

\bibitem{Dubnicka:2011mm}
  S.~Dubnicka, A.~Z.~Dubnickova, M.~A.~Ivanov, J.~G.~Koerner, P.~Santorelli,
  and G.~G.~Saidullaeva,
Phys. Rev. D \textbf{84}, 014006 (2011)
[arXiv:1104.3974 [hep-ph]].

\bibitem{Gutsche:2012ze}
T.~Gutsche, M.~A.~Ivanov, J.~G.~K\"orner, V.~E.~Lyubovitskij, and P.~Santorelli,
Phys. Rev. D \textbf{86}, 074013 (2012)
[arXiv:1207.7052 [hep-ph]].


\bibitem{Dubnicka:2024geu}
  S.~Dubni\v{c}ka, A.~Z.~Dubni\v{c}kov\'a, M.~A.~Ivanov, A.~Liptaj,
  A.~Tyulemissova, and Z.~Tyulemissov,
Phys. Rev. D \textbf{110}, 056030 (2024)
[arXiv:2406.09763 [hep-ph]].


\bibitem{Ivanov:2005fd}
M.~A.~Ivanov, J.~G.~K\"orner, and P.~Santorelli,
Phys. Rev. D \textbf{71}, 094006 (2005)
[erratum: Phys. Rev. D \textbf{75}, 019901 (2007)]
[arXiv:hep-ph/0501051 [hep-ph]].

\bibitem{Tran:2023hrn}
C.~T.~Tran, M.~A.~Ivanov, P.~Santorelli, and Q.~C.~Vo,
Chin. Phys. C \textbf{48}, 023103 (2024)
[arXiv:2311.15248 [hep-ph]].

\bibitem{Speagle:2019ivv}
J.~S.~Speagle,
Mon. Not. Roy. Astron. Soc. \textbf{493}, 3132(2020)
[arXiv:1904.02180 [astro-ph.IM]].

\bibitem{Virtanen:2019joe}
P.~Virtanen
\textit{et al.}
Nature Meth. \textbf{17}, 261 (2020)
[arXiv:1907.10121 [cs.MS]].

\bibitem{Wilks:1938dza}
S.~S.~Wilks,
Annals Math. Statist. \textbf{9}, no.1, 60 (1938).

\bibitem{Tran:2024phq}
C.~T.~Tran, M.~A.~Ivanov, P.~Santorelli, and H.~C.~Tran,
Chin. Phys. C \textbf{49},  013111 (2025)
[arXiv:2408.13776 [hep-ph]].

\bibitem{Tran:2025fhb}
C.~T.~Tran, M.~A.~Ivanov, and A.~T.~T.~Nguyen,
[arXiv:2506.23372 [hep-ph]].

\end{thebibliography}
